\documentclass[preprint,amsmath,amssymb]{revtex4-2}

\usepackage{graphicx}
\usepackage{dcolumn}
\usepackage{bm}
\usepackage{multirow}
\usepackage{float}
\usepackage{wrapfig}
\setcitestyle{super}
\usepackage[normalem]{ulem}
\usepackage{color}
\usepackage{booktabs}
\usepackage{longtable}
\usepackage{xr}
\usepackage{hyperref}
\usepackage{cleveref}
\usepackage{soul}
\usepackage[dvipsnames]{xcolor}

\makeatletter
\providecommand \@ifxundefined [1]{%
 \@ifx{#1\undefined}
}%
\providecommand \@ifnum [1]{%
 \ifnum #1\expandafter \@firstoftwo
 \else \expandafter \@secondoftwo
 \fi
}%
\providecommand \@ifx [1]{%
 \ifx #1\expandafter \@firstoftwo
 \else \expandafter \@secondoftwo
 \fi
}%
\providecommand \natexlab [1]{#1}%
\providecommand \bibnamefont  [1]{#1}%
\providecommand \bibfnamefont [1]{#1}%
\providecommand \citenamefont [1]{#1}%
\providecommand \href@noop [0]{\@secondoftwo}%
\providecommand \href [0]{\begingroup \@sanitize@url \@href}%
\providecommand \@href[1]{\@@startlink{#1}\@@href}%
\providecommand \@@href[1]{\endgroup#1\@@endlink}%
\providecommand \@sanitize@url [0]{\catcode `\\12\catcode `\$12\catcode
  `\&12\catcode `\#12\catcode `\^12\catcode `\_12\catcode `\%12\relax}%
\providecommand \@@startlink[1]{}%
\providecommand \@@endlink[0]{}%
\providecommand \url  [0]{\begingroup\@sanitize@url \@url }%
\providecommand \@url [1]{\endgroup\@href {#1}{\urlprefix }}%
\providecommand \urlprefix  [0]{URL }%
\providecommand \selectlanguage [0]{\@gobble}%
\providecommand \bibinfo  [0]{\@secondoftwo}%
\providecommand \bibfield  [0]{\@secondoftwo}%
\providecommand \BibitemOpen [0]{}%
\providecommand \BibitemShut  [1]{\csname bibitem#1\endcsname}%
\let\auto@bib@innerbib\@empty

\begin{document}
	
	\preprint{}
	\title{Orbital fluctuations and spin-orbital-lattice coupling in Bi$_2$Fe$_4$O$_9$}
	
	\author{Aditya Prasad Roy$^{1}$}
	\author{M. K. Chattopadhyay$^{2,3}$}
	\author{Ranjan Mittal$^{3,4}$}
	\author{Srungarpu N. Achary$^{3,5}$}
	\author{Avesh K. Tyagi$^{3}$}
	\author{Manh Duc Le$^{6}$}
	\author{Dipanshu Bansal$^{1}$}
	
	\affiliation{$^1$Indian Institute of Technology Bombay, Mumbai, MH 400076, India}
	\affiliation{$^2$Free Electron Laser Utilization Laboratory, Raja Ramanna Centre for Advanced Technology, Indore MP 452013, India}
	\affiliation{$^3$Homi Bhabha National Institute, Training School Complex, Anushakti Nagar, Mumbai MH 400094, India}
	\affiliation{$^4$Solid State Physics Division, Bhabha Atomic Research Centre, Mumbai, MH 400085, India}
	\affiliation{$^5$Chemistry Division, Bhabha Atomic Research Centre, Trombay, Mumbai 400085,	India}
	\affiliation{$^6$ISIS Neutron and Muon Source, Rutherford Appleton Laboratory, Chilton, Didcot, OX11 0QX, Oxfordshire, UK}
	\date{\today}
	
	\begin{abstract}
	
	    	Magnetic frustrations and degeneracies profoundly affect ground-state magnetic properties emerging from competing exchange interactions. Controlling such frustrations using orbital and phonon engineering via the Kugel-Khomskii-type (KK-type) interactions has recently enabled the orbital enhancement of magnetoelectric (ME) coupling. Using combined spectroscopic techniques and first-principle simulations, here we demonstrate that the magnetically frustrated Cairo lattice, Bi$_2$Fe$_4$O$_9$, exhibits a strong KK-type interaction resulting in a coupled spin-orbital phase below 1.8 times the N\'eel temperature ($T_{\rm N}\sim$245\,K). We observe an order of magnitude change in phonon linewidths that is not explainable considering spin-phonon coupling channels alone. Instead, the observed change is reminiscent of orbitally active materials, which we explicitly confirm by measuring the $T-$dependence of low-energy orbital excitations. We further find that Bi$_2$Fe$_4$O$_9$ harbors an unstable polar mode, driving the lattice to a symmetry-lowered ferroelectric (FE) phase below $T_{\rm N}$, in line with the previously reported hysteresis in polarization. Nonetheless, the FE phase leads to extremely small calculated superlattice Bragg peak intensities that are yet to be experimentally confirmed. Moreover, thermal conductivity measurements do not show any measurable effect of KK-type interactions on thermal transport across $T_{\rm N}$. But, we observe a repeatable anomaly at $\sim$57\,K appearing only in the heating cycle, which co-occurs with the $\sim$400\,meV broad continuum observed in Raman measurements. The observed KK-type interaction in Bi$_2$Fe$_4$O$_9$ provides an opportunity for orbital enhancement of ME coupling by phonon control of superexchange interactions.

	\end{abstract}
	
	\maketitle
	
	\pagebreak
	\pagebreak
	
	\section{Introduction}
	As pointed out by Wannier in the 1950s,~\cite{wannier1950antiferromagnetism} a two-dimensional (2D) triangular lattice of collinear antiferromagnetic (AFM) spin arrangement is geometrically frustrated, and, thereby, possesses multiple degenerate ground states. To lift the degeneracy and lower the system energy, frustrated spins reorient below the spin ordering temperature ($T_{\rm C}$) and are no longer antiparallel. In certain cases, reorientation via canting or non-collinear spin ordering breaks the crystallographic inversion symmetry and drives the lattice toward its non-centrosymmetric polar state. An entwined frustrated spin and lattice system can lead to an anomalous response of physical observables, such as the pyroelectric effect across $T_{\rm C}$ in Cr$_2$BeO$_4$.~\cite{newnham1978magnetoferroelectricity} These entwined systems have also led to the discovery of novel responses, including the magnetoelectric (ME) effect.~\cite{fiebig2016evolution,arima2006collinear}
	
	\par Non-collinear spin-order induced ME state has one broadly discussed origin, the inverse Dzyaloshinskii–Moriya (DM) interaction.~\cite{katsura2005spin, mostovoy2006ferroelectricity} Here, an antisymmetric exchange of neighboring intersite spins due to relativistic spin-orbit coupling expressed as ${\bf S}_i \times {\bf S}_j$, promotes inversion symmetry breaking displacement of ligand ions. Since in a relativistic spin-orbit coupling, magnetic ions with non-zero orbital angular momentum couple to the spins, it leads to an extra degree of freedom between the spin and lattice. Nevertheless, owing to the relativistic nature, induced polarization ($\mathbf{P}$) remains weak,~\cite{kocsis2023spin} with values of the order of $\mathbf{P}\sim$0.1~$\,\mu$Ccm$^{-2}$; for example, $\mathbf{P}\sim$0.3~$\,\mu$Ccm$^{-2}$ in CaMn$_7$O$_{12}$.~\cite{johnson2012giant}
	
	\par In an alternate mechanism distinct from the relativistic coupling, spin-orbit interaction occurs via spin-induced orbital polarization.~\cite{kugel1973crystal,khomskii2003orbital,roy2024evidence,disa2023photo,lee2012two} Here, orbital orientation gets confined to the spin order, leading to the Kugel-Khomskii-type (KK) spin-orbital coupling.~\cite{kocsis2023spin} Recent Raman measurements on rare-earth titanates, vanadates, and Swedenborgites,~\cite{keimer2000spin, ulrich2006raman,ulrich2009momentum, ulrich2015spin, miyasaka2005one, sugai2006orbital, kocsis2023spin} reveal that above $T_{\rm C}$, the presence of low-energy collective orbital fluctuations (OFs) directly couple to phonons via chemical bond modulation and strongly scatter them. However, below $T_{\rm C}$, the spin-ordered state suppresses the OFs and their collective low-energy excitations due to orbital confinement. Such ordering of the orbital states confined to spins explains the change in observed phonon linewidths by nearly an order of magnitude across $T_{\rm C}$ in the above materials. In some cases, the orbitals could also be confined by short-range magnetic ordering and show strong orbital-phonon scattering well above the transition temperature. For example, recently investigated CrVO$_4$ revealed a strong interplay between spin, orbital, and lattice degrees via the KK-type interaction,~\cite{roy2024evidence} harboring anomalies in physical observables persisting up to four times its $T_{\rm C}$. Thus, spin-orbit interaction occurring via KK-type interaction could provide a potential pathway to explore orbital enhancement of ME coupling, i.e., spin-induced $\mathbf{P}$. Indeed, measurements of phonons and dielectric response in CaBa$M_4$O$_7$ ($M = $Co, Fe), possessing orbital degeneracy and non-degeneracy depending on the composition, have provided the first direct evidence of orbital enhancement of ME coupling.~\cite{kocsis2023spin}
			
	\par Here, we investigate the role of orbital degrees in the Cairo lattice Bi$_2$Fe$_4$O$_9$ having a N\`eel temperature $T_{\rm N} \sim$ 245\,K~\cite{ressouche2009magnetic} that results in significant changes in physical observables. We find an order of magnitude change in phonon linewidths ($\Delta\Gamma$) that can not be accounted for using the spin-phonon coupling channel alone. Instead, the observed behavior is reminiscent of orbitally active materials, LaTiO$_3$,~\cite{ulrich2015spin} YTiO$_3$,~\cite{disa2023photo} CaBa(Co, Fe)$_4$O$_7$,~\cite{kocsis2023spin} and CrVO$_4$.~\cite{roy2024evidence} The orbital degree, originating from partially filled Fe$^{3+}$ $d-$orbitals, is confirmed by directly measuring low-energy orbital excitations, and their $T-$dependence highlights strong KK-type interactions below 1.8$T_{\rm N}$, i.e., a coupled spin-orbital phase. We further find that all modes that show anomalous behavior perturb the Fe$_{\rm oct}$-O$_{\rm ap}$-Bi-O$_{\rm ap}$-Fe$_{\rm oct}$ exchange pathway, thus highlighting a direct interaction between orbitals contributing to exchange integral and phonons. Moreover, the relatively small change in phonon energy due to long-range magnetic order below $T_{\rm N}$ shows that Bi$_2$Fe$_4$O$_9$ harbors weak spin-phonon coupling. Interestingly, we find an unstable polar mode involving the Bi off-centering motion along the $c$-axis, which induces a measurable electronic $\mathbf{P}$. The calculated $\mathbf{P}$ values from polar mode condensation are comparable to previously reported $\mathbf{P}$.~\cite{singh2008substantial} Furthermore, we observe a hitherto unreported distinct anomaly in thermal conductivity ($\kappa$) measurements at $\sim$57\,K on the heating cycle, which is correlated with a large renormalization at $\sim$400\,meV in Raman measurements. However, further theoretical and experimental studies are necessary for deciphering its origin.

	\section{Methods}
	\subsection{Sample synthesis}\label{Sample preparation}
	\par Using the solid-state reaction method, homogeneously mixed pellets of stoichiometric Bi$_2$O$_3$ (Merck) and Fe$_2$O$_3$ (Merck) were heated at 750\,$^{\circ}$C for 15\,h on a platinum foil. Thereafter, the pellets were crushed, regrounded, and reheated at 850\,$^{\circ}$C for calcination. Following this, the samples were regrounded, repressed into pellets, and heated at 850\,$^{\circ}$C for 15\ h. At this stage, we observed the formation of a single-phase Bi$_2$Fe$_4$O$_9$. Further, the obtained sample is regrounded and sintered at 900\,$^{\circ}$C for 10\,h. The characterization of the sample with X-ray diffraction is discussed below.

	\subsection{X-ray diffraction}\label{X-ray diffraction}
	The Bi$_2$Fe$_4$O$_9$ polycrystalline sample obtained from solid-state reaction method was characterized by x-ray diffraction (XRD) using Rigaku instrument with Cu-K$\alpha_1$ ($\lambda$ = 1.54060\,\AA) source and D/tex ultra 1D detector providing resolution of 2$\theta$ = 0.01$^{\circ}$ in $\theta/2\theta$ scanning mode. Diffraction patterns were scanned between 10-110$^{\circ}$ at 3$^{\circ}$ per minute after placing the powder sample on a glass slide. The obtained XRD patterns were refined using GSAS-II software.~\cite{toby2013gsas} The crystal structure refinement at ambient is shown in Supplementary Materials (SM) Fig.~S1.~\cite{SM_ref_Bi2Fe4O9} The resulting centrosymmetric orthorhombic structure ($Pbam$, space group no. 55), with $a =$ 7.969\,\AA, $b =$ 8.435\,\AA, and $c =$ 5.998\,\AA, $V$ = 403.274\,\AA$^3$ is consistent with previous reports.~\cite{shamir1978magnetic}
	
	\subsection{Spectroscopy measurements}
	\subsubsection{Inelastic neutron scattering measurements}\label{Inelastic neutron scattering measurements}
	Inelastic neutron scattering (INS) measurements were performed at the MARI time-of-flight chopper spectrometer, ISIS Neutron and Muon Source, UK. About $\sim$5\,g powder sample was encased inside a thin-walled aluminum can. A second aluminum can in empty condition was used to measure the background signal. All the $T-$dependent measurements were performed using a closed-cycle helium cryostat. Incident neutron energies ($E_i$) of 27, 80, and 140 meV allowed us to measure magnons and phonons across $T_{\rm N}$. The Fermi chopper was set to 400\,Hz for $E_i =$ 27 and 140\,meV and 300\,Hz for $E_i$ = 80\,meV, respectively. The energy resolution (full width at half maximum) for $E_i = 140$\,meV was $\sim$4.7\,meV at the elastic line, which reduces to $\sim$3\,meV at an energy transfer of 60\,meV. Similarly, the resolution for $E_i = 80$ and 27\,meV was $\sim$2.4 and $\sim$0.5\,meV at the elastic line. The obtained signal was transformed from instrument coordinates, scattering angle, and neutron detection times to physical momentum transfer ($Q$) and energy transfer ($E$) using MANTID software.~\cite{arnold2014mantid} The transformed coordinates were further used to find two-dimensional maps of dynamical structure factor S(Q, $E$), and subsequently the dynamical susceptibility, $\chi''(Q, E)$, from $S(Q, E)$ using $\chi''(Q, E) = (1-\exp(-E/k_{\rm B}T)) \ S(Q, E)$. Here, $k_{\rm B}$ is the Boltzmann constant. We obtain phonon density of states (PDOS) by integrating over 5 $\leqslant$ $Q$ $\leqslant$ 11 \AA$^{-1}$ (magnetic intensities have small contributions above 5\,\AA$^{-1}$), and applying background, multiphonon, and multiple scattering corrections as described in our earlier study.~\cite{bansal2015electron} The phonon spectra from INS measurements are weighted by the relative neutron scattering strengths (${\sigma}/{m}$) of individual elements, where ${\sigma}$ is the total neutron scattering cross-section and $m$ is the atomic mass. Neutron-weighting (NW) factors for Bi, Fe, and O are ${\sigma_{Bi}}/{m_{Bi}} = 0.0438$, ${\sigma_{Fe}}/{m_{Fe}} = 0.2224$, and ${\sigma_{O}}/{m_{O}} = 0.2645$, respectively (in units of barns/amu). Consequently, the NW-PDOS, $g_{\rm NW}(E)$, is expressed as: 
	\begin{eqnarray}
		g_{NW}(E) &=& \left( \frac{\sigma_{Bi}}{m_{Bi}} g_{Bi}(E) +  \frac{\sigma_{Fe}}{m_{Fe}} g_{Fe}(E) + \frac{\sigma_{O}}{m_{O}} g_{O}(E)\right) /\left( \frac{\sigma_{Bi}}{m_{Bi}} +  \frac{\sigma_{Fe}}{m_{Fe}} +  \frac{\sigma_{O}}{m_{O}} \right)
	\end{eqnarray}
	\noindent Here, $g_{\rm Bi}(E)$, $g_{\rm Fe}(E)$, $g_{\rm O}(E)$ are partial PDOS of Bi, Fe, and O, respectively. The resulting NW-PDOS is normalized to its unit area and can be directly compared against the simulated PDOS after applying the same NW to the simulations.
	
	\subsubsection{Raman measurements}\label{Raman spectroscopy}
	Raman measurements in backscattering geometry were performed using a Renishaw inVia$^{\rm TM}$ confocal Raman spectrometer equipped with a Peltier-cooled charge-coupled device detector. We used a 532\,nm diode pumped solid state laser as the excitation source and subsequently focused on the sample using a 50X objective with a laser power of $<$2.5\,mW to avoid local heating. A diffraction grating of 2400\,grooves/mm provided a spectral resolution of $<$1\,cm$^{-1}$. The spectrometer was calibrated using the 521\,cm$^{-1}$ mode of Si thin film. Lakeshore model number 335 with He-exchange-based cryostat (with vaccum chamber providing $\sim10^{-6}$\,mbar) and Linkam stage TS1500 temperature controllers were used for $T$-dependent measurements spanning 15 to 800\,K (stability of $\pm$ 0.1\,K). Measurements were taken from 50-4000\,cm$^{-1}$. The measured spectra were fitted using Lorentzian peaks to extract phonon energies ($\omega$) and line widths ($\Gamma$) at full-width half maxima (FWHM). 
	
	\par Photon-wavelength-dependent photoluminescence (PL) measurements were performed using two excitation sources at 532 and 633\,nm, and 600 and 1800\,grooves/mm diffraction gratings, allowing to cover a wide range spanning up to $\sim$10000 and 8000\,cm$^{-1}$, respectively, using the same Raman spectrometer. 
		
	\subsection{Heat capacity}\label{sec_Cp}
	A two-tau relaxation calorimetry was used to perform $C_p$ measurements on the Dynacool Physical Property Measurement System (PPMS, Quantum Design). A small chunk of mass $\sim$20\,mg was cut from the parent sintered pellet and mounted on the platform using Apiezon N grease. Measurements were performed between 2-300\,K.
	
	\subsection{Thermal Conductivity}\label{sec_kappa}
	For thermal conductivity ($\kappa$) measurements, we used the thermal transport option of liquid He-based 9\,T PPMS (Quantum Design). A rectangular rod
	of dimensions $\sim$ 13.4×3.75×2.43\,mm$^{3}$ was retrieved from the parent sintered pellet for doing $\kappa$ measurements. Four gold-coated copper probes were attached to this rod using indium solder. The temperature sensors, heater, and ground connections were attached to these probes using gold-coated copper screws.
	$\kappa$ measurements spanning 2-350\,K were performed in 0 and 8\,T magnetic fields.
	
	\subsection{Simulations}
	\subsubsection{Phonon simulations}\label{Harmonic phonon simulation}
	We perform electronic structure calculations within the density functional theory (DFT) framework as implemented in the Vienna ab initio Simulation Package (VASP).~\cite{kresse1993ab,kresse1996efficient,kresse1996efficiency} We use an 8$\times$6$\times$6 $\Gamma$-centered Monkhorst-Pack electronic $k$-point mesh to integrate over the entire Brillouin zone with a converged plane-wave cut-off energy ($E_{Gcut}$) of 650\,eV on a 30-atom basis unit cell ($<$1\,meV/atom for $E_{Gcut}$ and $k$-mesh). The convergence criteria were set to $10^{-8}$ eV for a self-consistent electronic loop. The projector-augmented-wave (PAW) potentials explicitly include five valence electrons for Bi ($6s^2 6p^3$), eight for Fe ($3d^7 4s^1$), and six for O ($2s^2 2p^4$). The generalized-gradient-approximation (GGA) within the Perdew-Burke-Ernzerhof description for solids (PBEsol)~\cite{csonka2009assessing} were used within the considered pseudopotentials. Furthermore, the GGA$+U$ calculations with on-site Coulomb interaction, $U = 4.2$\,eV, following the method of Dudarev \emph{et al.}~\cite{dudarev1998electron} reproduces the observed bandgap obtained in optical measurements ($\sim$2.1\,eV~\cite{irshad2015first}). To limit computational expenses for phonon dispersions, the observed non-colinear AFM order (refer to main text Fig.1(c)) was projected on Fe$_{\rm tet}$ and Fe$_{\rm oct}$ sites along the $a-$axis for the spin-polarized calculations. The relativistic spin-orbit coupling was considered for initial testing and was not included subsequently. During the relaxation step, lattice parameters and ionic positions were optimized until forces on individual atoms were reduced below 1\,meV\AA$^{-1}$. The resulting relaxed lattice parameters ($a =$ 7.886\,\AA, $b =$ 8.459\,\AA, and $c =$ 6.012\,\AA) agree reasonably well with the experimental values. 
	
	\par Finite displacement approach within the harmonic approximation as implemented in the Phonopy code~\cite{togo2015first} were used to calculate phonon dispersions. We used a 2$\times$2$\times$3 supercell of a primitive cell containing 360 atoms. We used 27 independent atomic displacements and perturbation amplitude of 0.01\,\AA~to calculate the interatomic force constants. Raman and Infrared (IR) activities were simulated using the Phonopy-Spectroscopy code.~\cite{skelton2017lattice} To calculate the Gr\"{u}neisen parameters at the $\Gamma-$point, we apply a 1\,$\%$ volumetric compression and expansion and find $\Delta \omega$ using the Phonopy code.
	
	\subsubsection{Spin-wave simulations}\label{Spin-wave simulations}
	Magnon dispersions were simulated using linear spin wave theory (LSWT), as implemented in SpinW.~\cite{toth2015linear} Experimentally observed non-colinear AFM order~\cite{shamir1978magnetic} (see main text Fig.~1(c)) with eight Fe atoms in the conventional cell and full moments on Fe$^{3+}$ ions were used. Following Le et al.,~\cite{le2021experimental}, we consider neighboring interactions up to 3.59\,\AA\ and the fitted exchange constants (as shown in main text Fig.~1(c)) $J_{33}$, $J_{43}$, $J_{43}^{\prime}$, $J_c$, and $J_{44}$ equal to 27.6, 3.1, 6.5, 1.39, and -0.22\,meV ($+$ve sign is for the AFM interaction). Further, the observed spin gap $\sim$3\,meV at 5\,K is reproduced in LSWT simulations by considering a single-ion-anisotropy term in the magnetic Hamiltonian that constrains Fe$^{3+}$ moments within the $a-b$ plane. Subsequently, powder averaging of the calculated dispersion was performed using a magnetic form factor, $S$ = 3/2, and nRand = 1000 random samples per Q-shell. A comparison of simulation with INS measurement at 5\,K (see main text Fig.~2(a) and (b)) shows an excellent agreement. 
	
	\section{Results and Discussion}
	\subsection{Phase Transition}
\begin{figure}[H]
		\begin{center}
			\includegraphics[trim=4.5cm 7.15cm 4.4cm 6.95cm, clip=true, width=0.65\textwidth]{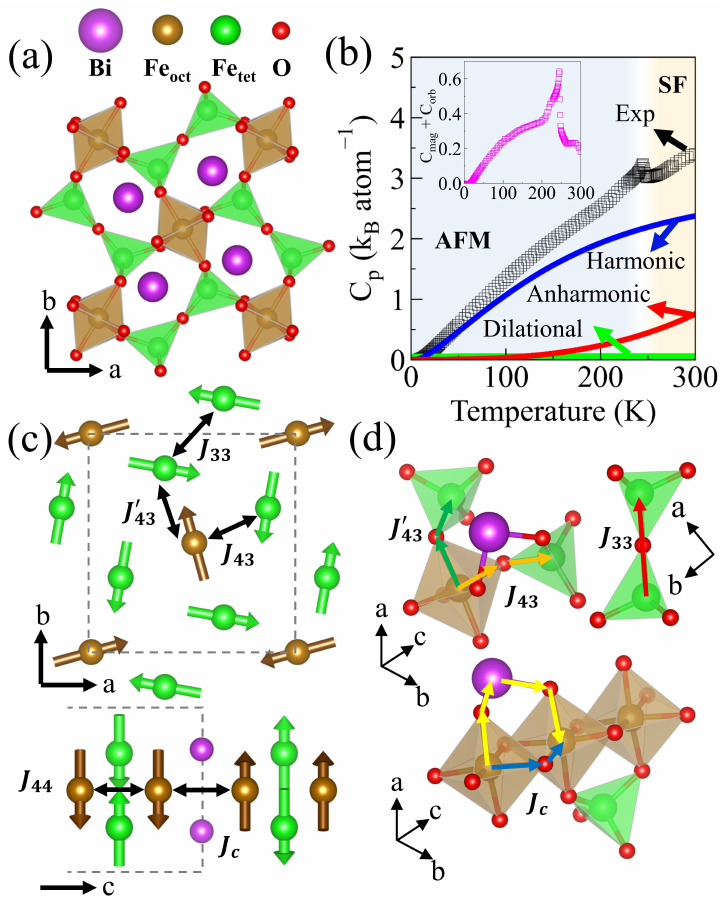}
		\end{center}
		\vspace{-0.2in}
		\caption[Crystal and magnetic structure, and heat capacity across $T_{\rm N}$.]{
			The Cairo pentagonal structure of Bi$_2$Fe$_4$O$_9$ projected along $a-b$ plane. Fe$_{\rm oct}$ and Fe$_{\rm tet}$ are shown with brown and green color polyhedron, respectively. Bi$^{3+}$ ions (purple color atoms) fill the pentagonal voids. (b) $C_p$ data between 2 to 300\,K showing a transition at $T_{\rm N}\sim$ 245\,K. Calculated harmonic, anharmonic, and dilational contributions are shown underneath the experimental data. The sum of magnetic and orbital contributions ($C_{mag}$ + $C_{orb}$) obtained by subtracting harmonic, anharmonic, and dilational contributions from the measured data is shown in the inset. Harmonic and dilational contributions are calculated using DFT simulations and thermal expansion from Ref~\citenum{krenzel2012thermal}. (c) The frustrated AFM structure projected along the $a-b$ plane and $c$-axis showing the dominant exchange interactions $J_{33}$, $J_{43}$, $J^{\prime}_{43}$, $J_{c}$, and $J_{44}$ between the magnetic Fe$^{3+}$ ions (see black arrows). (d) Color scheme of various superexchange pathways -- (i) Fe$_{\rm oct}$-O$_{\rm ap}$-Fe$_{\rm tet}$ (green), (ii) Fe$_{\rm oct}$-O$_{\rm eq}$-Fe$_{\rm tet}$ (orange), (iii) Fe$_{\rm tet}$-O-Fe$_{\rm tet}$ (red), and (iv) Fe$_{\rm oct}$-O$_{\rm ap}$-Bi-O$_{\rm ap}$-Fe$_{\rm oct}$ (yellow) and Fe$_{\rm oct}$-O$_{\rm eq}$-Fe$_{\rm oct}$ (blue), contribute to $J_{43}^{\prime}$, $J_{43}$, $J_{33}$, and $J_{c}$, respectively.
		}
		\label{fig:Crystal_Structure}
	\end{figure}

	Bi$_2$Fe$_4$O$_9$ crystallizes in the orthorhombic structure (see  Fig.~\ref{fig:Crystal_Structure}(a)) at room temperature, having a $Pbam$ space group (No. 55). The Cairo lattice composes of Fe-sites distributed within edge-shared FeO$_6$ octahedra (Fe$_{\rm oct}$) and corner-shared FeO$_4$ tetrahedra (Fe$_{\rm tet}$). Bi$^{3+}$ occupies the pentagonal voids between two polyhedrons, alternating along the $c-$axis. Our X-ray diffraction (XRD) measurements on the powder sample (see SM Fig.~S1) confirm the structure. Below $T_{\rm N} \sim$ 245\,K (refer to Fig.~\ref{fig:Crystal_Structure}(b) for $C_p$ measurements), Bi$_2$Fe$_4$O$_9$ undergoes an AFM phase transition characterized by a propagation vector of $\mathbf{k}$=($\frac{1}{2},\frac{1}{2},\frac{1}{2}$)  r.l.u.,~\cite{shamir1978magnetic} implying that the magnetic unit cell is doubled along all three crystallographic axes ([a$_{\rm m}$,b$_{\rm m}$,c$_{\rm m}$] = 2[a,b,c]). Here, r.l.u.~refers to the reciprocal lattice units. In the AFM phase, Fe$^{3+}$ moments are parallel to the $a-b$ plane with a finite canting, such that moments on Fe$_{\rm tet}$ and Fe$_{\rm oct}$ sites subtend a global rotation of $\sim$155$^{\circ}$ (see Fig.~\ref{fig:Crystal_Structure}(c)).~\cite{ressouche2009magnetic,beauvois2020dimer} Additionally, the constrained moments exhibit AFM order across the interlayer Bi$^{3+}$ ion. Although already reported in the literature,~\cite{ressouche2009magnetic,beauvois2020dimer} the magnetic ordering is confirmed by the appearance of magnetic Bragg peaks below $T_{\rm N}$ in our elastic neutron scattering measurements (see SM Fig.~S2).
	
	\subsection{Exchange Interactions and Magnons}\label{EIM}
\begin{figure}[H]
		\begin{center}
			\includegraphics[trim=5.5cm 3.25cm 7.6cm 3.8cm, clip=true, width=0.7\textwidth]{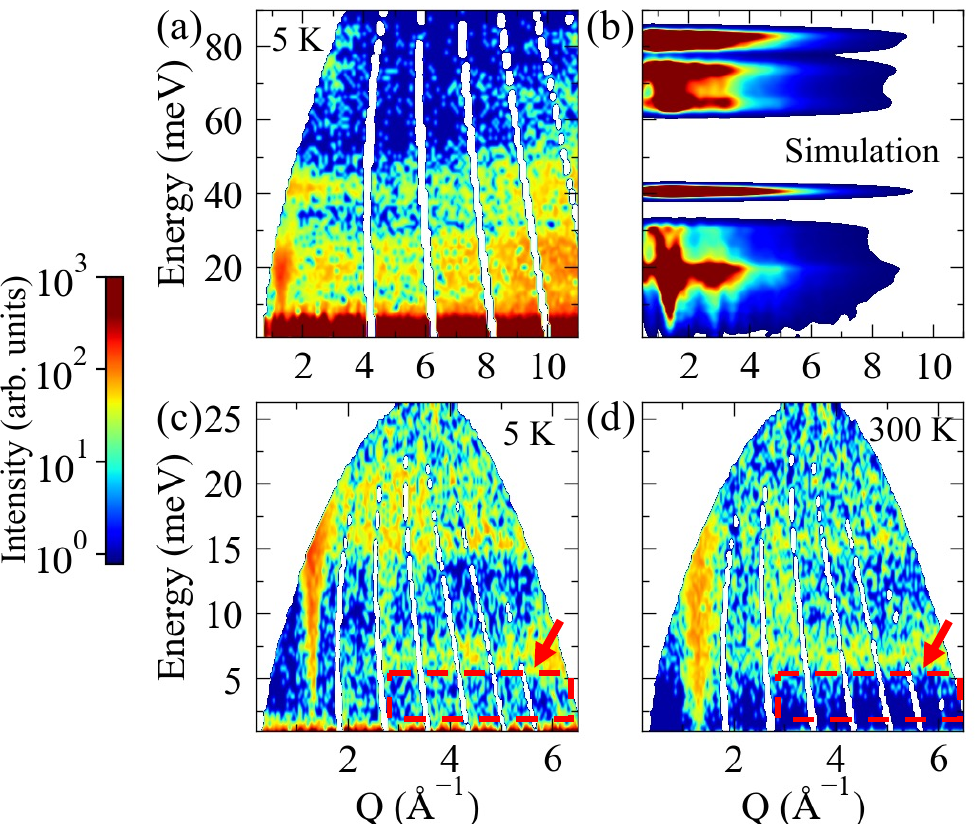}
		\end{center}
		\caption[The magnon dispersion.]{
			$\chi^{\prime \prime}(Q,E)$ measured using INS with (a) $E_i$ = 140\,meV showing dispersing magnons from magnetic Bragg peaks at low-$Q$. Phonon intensity is visible at high-$Q$, i.e., $Q>6$\,\AA$^{-1}$. (b) The simulated magnon $\chi^{\prime \prime}(Q,E)$ from LSWT calculations using the exchange interactions described in the text. (c, d) Same as panel (a), expect measurements are for $E_i$ = 27\,meV at 5 and 300\,K. A magnon gap is clearly visible near $Q = 1.3$\,\AA$^{-1}$. Panel (d) shows a blob of magnetic intensity at 300\,K, thus confirming the short-range spin fluctuations (SRSF) above $T_{\rm N}$. Dotted red boxes and arrows in panels (c) and (d) show the appearance of low energy intensities ($\sim$3\,meV for Q$>$3\,\AA$^{-1}$) arising from phonon contributions. 
		}
		\label{fig:Spin_Wave}
	\end{figure}

	The complex AFM structure emerges from five dominant exchange interactions between magnetic Fe$^{3+}$ ions, as identified in recent single-crystal INS measurements.~\cite{beauvois2020dimer,le2021experimental} We use the labeling of exchange interactions as adopted in Ref.~\citenum{le2021experimental}. Figure~\ref{fig:Crystal_Structure}(c) highlights various exchange interactions. Fe$_{\rm tet}$ sites interact via $J_{33}$ through corner shared O atom. The inplane $J_{43}$ and $J_{43}^{\prime}$ form the exchange interactions between Fe$_{\rm oct}$ and Fe$_{\rm tet}$ sites. The out-of-plane $J_{44}$ and $J_{c}$ lie between the intralayer and interlayer Fe$_{\rm oct}$-sites, respectively, along the $c-$axis. The resulting ground-state magnetic structure emerging from the complex magnetic interactions leads to intriguing magnon dispersion.~\cite{beauvois2020dimer,le2021experimental} We probe the magnons using INS measurements in a powder sample across $T_{\rm N}$ using three incident energies to both get the high-resolution data, cover the entire magnon spectrum, and conclusively separate the magnon signal from phonons. Figure~\ref{fig:Spin_Wave}(a) shows the dynamical susceptibility ($\chi^{\prime\prime}(Q, E)$) measured at 5\,K, highlighting dispersing magnons emerging from the magnetic Bragg peaks. Here, we differentiate between magnons and phonons from their $Q$-dependence. While magnon intensity decays with increasing $Q$, phonon intensity increases as $Q^{2}$. Data shows dispersing magnons rising from magnetic Bragg peaks and extending to 50\,meV for $Q<$6\,\AA$^{-1}$. Further, consistent with Ref.~\citenum{le2021experimental}, we also observe a finite magnon intensity between 60 to 90\,meV. Figure~\ref{fig:Spin_Wave}(a) shows that the magnon intensity between 60 to 90\,meV diminishes for $Q>$5\,\AA$^{-1}$, whereas the phonon intensity lying in the similar energy range increases above $Q>$9\,\AA$^{-1}$. The simultaneous measurement of phonons and magnons allows us to conclusively assign the signal below $Q\sim$6\,\AA$^{-1}$ to magnons. 
		
	\par The obtained magnon dispersion enables a comprehensive understanding of various magnetic exchange pathways via simulations. We reproduce the dispersion using linear spin wave theory (LSWT) and refined $J$'s from Ref.~\citenum{le2021experimental} (see Fig.~\ref{fig:Spin_Wave}(b)). $J$ values are listed in Methods section~\ref{Spin-wave simulations}. A largest $J_{33}$ = 27\,meV occurs via 180$^{\circ}$ Fe$_{\rm tet}$-O-Fe$_{\rm tet}$ superexchange interaction (see Fig.~\ref{fig:Crystal_Structure}(d) for various superexchange pathways discussed here), consistent with the Goodenough-Kanamori-Anderson rule. Both $J_{43}$ and $J_{43}^{\prime}$ exchange interactions between Fe$_{\rm oct}$ and Fe$_{\rm tet}$ sites share a common Fe$_{\rm oct}$-O-Fe$_{\rm tet}$ superexchange pathway (O$_{\rm eq}$ and O$_{\rm ap}$ for $J_{43}$ and $J_{43}^{\prime}$ at 121$^{\circ}$ and 131$^{\circ}$, respectively). Here, O$_{\rm eq}$ and O$_{\rm ap}$ refer to equatorial and apical O atoms in the FeO$_6$ octahedra. The near 120$^{\circ}$ superexchanges lie between a strong 180$^{\circ}$ AFM and weak 90$^{\circ}$ ferromagnetism (FM),~\cite{geertsma1996influence} and here results in a net AFM order. Interestingly, $J_{43}$ is twice of $J_{43}^{\prime}$, which, as elaborated in Ref.~\citenum{pchelkina2013ab}, is partially due to the enhancement in exchange via $t_{2g}/t_{2g}$ and $t_{2g}/e_{g}$ channels in $J_{43}$. Additionally, the intralayer $J_{c}$ and interlayer $J_{44}$ along the $c-$axis share a common near 90$^{\circ}$ Fe$_{\rm oct}$-O$_{\rm eq}$-Fe$_{\rm oct}$ superexchange pathway. As discussed for CuGeO$_3$,~\cite{geertsma1996influence} a Fe$_{\rm oct}$-O$_{\rm eq}$-Fe$_{\rm oct}$ path allows weak FM interaction of $J_{44}$ via (i) Fe$_{\rm oct}$ exchange with O$_{\rm eq}$ $p-$orbitals polarized by another Fe$_{\rm oct}$ site and via (ii) $p-p$ hybridization. In contrast, $J_c$, besides Fe$_{\rm oct}$-O$_{\rm eq}$-Fe$_{\rm oct}$, also has contribution from Fe$_{\rm oct}$-O$_{\rm ap}$-Bi-O$_{\rm ap}$-Fe$_{\rm oct}$ (see Fig.~\ref{fig:Crystal_Structure}(d)), which leads to a net dominant AFM interaction. Exchange interactions via Bi ion are evident from polarized neutron scattering measurements of Beauvois et al.~\cite{beauvois2020dimer}, where they reported a measurable magnetic density on Bi atoms. Nuclear Magnetic Resonance studies on a similar class of materials, Bi$_2$CuO$_4$,~\cite{gippius1998observation} also confirm the Bi atom contribution via a longer superexchange pathway. As described in Ref.~\citenum{beauvois2020dimer}, this weak $J_{c}$ triggers the long-range magnetic transition to the AFM phase. Furthermore, consistent with previous measurements,~\cite{le2021experimental} we also observe a gap at the zone-center corresponding to easy-plane single-ion anisotropy (see near $Q = 1.3$\,\AA$^{-1}$ in Fig.~\ref{fig:Spin_Wave}(c)). The above-discussed superexchange mechanisms in Bi$_2$Fe$_4$O$_9$ result in the observed non-collinear ground state AFM configuration. 
		
	\subsection{Kugel-Khomskii-type Interaction Induced Anomalous Phonon Behavior}\label{APSS}
	INS measurements of magnons above $T_{\rm N}$ (see Fig.~\ref{fig:Spin_Wave}(d)) revealed a blob of magnetic intensity at 300\,K, suggesting the presence of short-range spin fluctuations (SRSF). The intensity from SRSF persists up to 1.8$T_{\rm N}\sim$440\,K (not shown). Such SRSF can enhance or suppress phonon scattering owing to the generation of an inhomogeneous strain field (for example, YMnO$_3$)~\cite{sharma2004thermal} or orbital polarization via KK-type interaction (for example, CrVO$_4$).~\cite{roy2024evidence} This prompted us to probe the effect of SRSF on phonons in Bi$_2$Fe$_4$O$_9$. 

	\par
\begin{figure}[H]
		\begin{center}
			\includegraphics[trim=0.31cm 5.75cm 8.8cm 2.7cm, clip=true, width=0.7\textwidth]{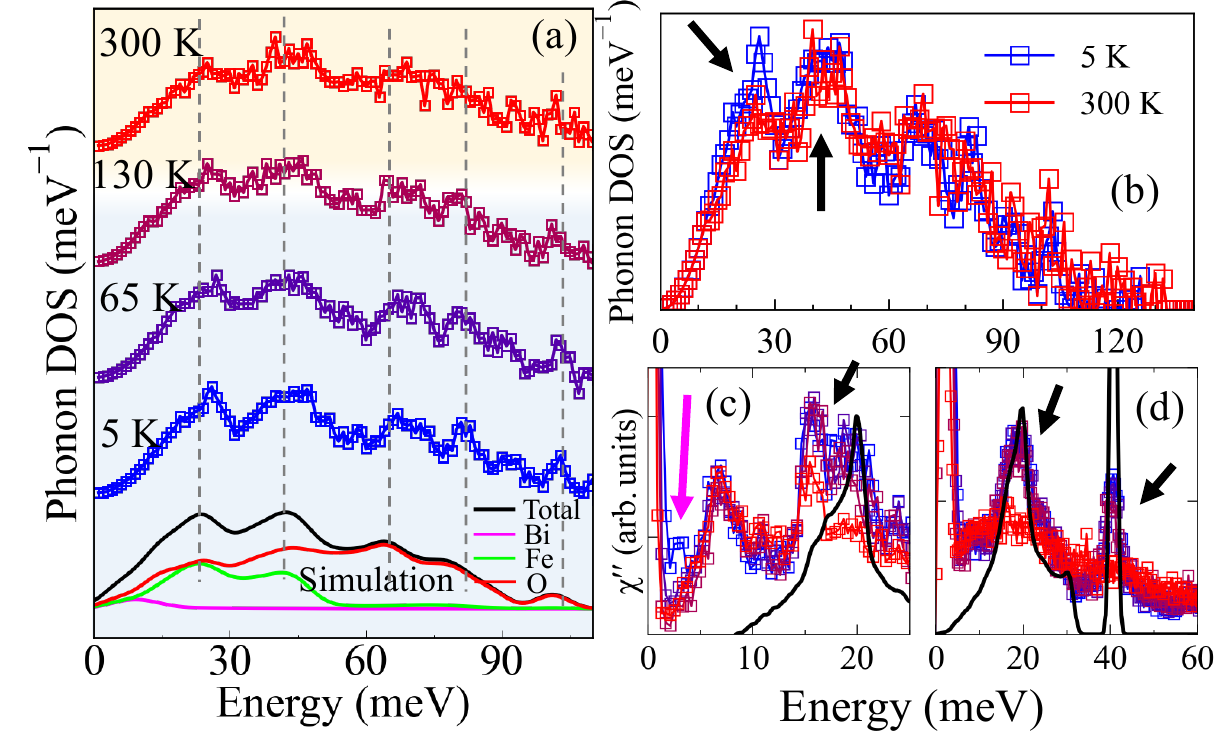}
		\end{center}
		\caption{
			(a) Comparison of neutron-weighted phonon density of states (NW-PDOS) of Bi$_2$Fe$_4$O$_9$ measured using INS across $T_{\rm N}$ ($E_i$ = 140 \,meV) with DFT simulated NW-PDOS. Individual NW atomic contributions to NW-PDOS are highlighted below the simulated PDOS. Curves are offset for clarity. (b) An Overplot of NW-PDOS data at 5 and 300\,K. The intensity change between two temperatures at approximately 20 and 40\,meV is highlighted by black arrows. The data here is the same as panel (a). (c,d) Comparison of $\chi''(E)$ at 5 and 300\,K measured using $E_i$ = 27 (panel c) and 80 \,meV (panel d), respectively. Black lines in panels (c) and (d) are integrated powder-averaged magnon intensity calculated using LSWT. From the measured and simulated curves at 5\,K, we confirm that the observed intensity reduction in panel (b) across $T_{\rm N}$ arises from magnon contributions (see black arrows). The magenta arrow in panel (c) highlights a gradually developing low energy ($\sim$3\,meV) peak arising from phonon contributions (see discussion in the text).
		}
		\label{fig:Phonon_DOS}
	\end{figure}

 First, we discuss the momentum-averaged phonon dynamics. Figure~\ref{fig:Spin_Wave}(a) shows INS measurement of magnons and phonons at 5\,K. Phonon intensities are visible for $Q>6$\,\AA$^{-1}$. By integrating the data over 5 $\leqslant$ $Q$ $\leqslant$ 11 \AA$^{-1}$ (magnetic intensities have small contributions above 5\,\AA$^{-1}$), and applying background, multiphonon, and multiple scattering corrections as described in our earlier study,~\cite{bansal2015electron} we calculate the phonon density of states (PDOS, see Fig.~\ref{fig:Phonon_DOS}(a)) across $T_{\rm N}$. On careful observation, we do not observe a measurable shift within our instrument resolution (see Methods section for resolution). However, phonon broadening is evident on heating owing to an increase in anharmonicity, as also evident from derived anharmonic $C_p$ (see Fig.~\ref{fig:Crystal_Structure}(d)). Figure~\ref{fig:Phonon_DOS}(b) overplots the PDOS at 5 and 300\,K, revealing the drop in intensity near 25 and 40\,meV. However, the drop is not from the phonon-magnon coupling but originates from the suppression of magnon intensity above $T_{\rm N}$.  Note that we integrated for $Q$ $\geqslant$ 5 \AA$^{-1}$ to avoid contribution from magnons, but due to non-zero contribution of magnons above $Q$ $\geqslant$ 5 \AA$^{-1}$, PDOS has small magnon contribution which we here observed as a drop in intensity. Figure~\ref{fig:Phonon_DOS}(c,d) overplots simulated magnon peaks over the measured data to explicitly confirm that the above drop across $T_{\rm N}$ is indeed from magnons. Notably, in the above data, we also observe a presence of finite intensity at $\sim$3\,meV at 5\,K (see pink arrow in panel c). This intensity was also visible at 130\,K, although with a much lower intensity. On careful analysis, we find this feature to originate from 3 $\leqslant$ $Q$ $\leqslant$ 7 \AA$^{-1}$, which disappears at 300\,K (see the red dashed rectangles and arrows in Fig.~\ref{fig:Spin_Wave}(c,d)). Since the intensity is prominently visible at higher $Q$, it probably originates from phonons. The same intensity feature was also captured in earlier INS measurements,~\cite{le2021experimental} and from longitudinal polarization analysis, was attributed to have a non-magnetic origin. Later in the text, we will interpret this feature as a low-energy Bi-dominated phonon excitation emerging from a lattice instability below $T_{\rm N}$.

\begin{figure}[H]
		\begin{center}
			\includegraphics[trim=7cm 4.15cm 13.6cm 2.95cm, clip=true, width=0.75\textwidth]{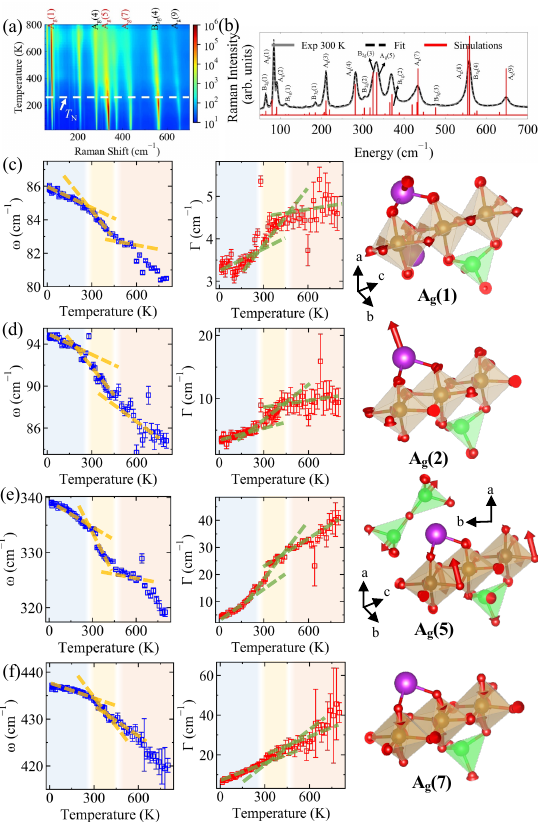}
		\end{center}
		\vspace{-0.35in}
		\caption[Phonons across $T_{\rm N}$.]{
			 (a) Color plot of Raman intensity from 15 to 800\,K. Selected phonons are labeled on top of the panel. Color bar denotes Raman intensity in arbitrary units. (b) Identification of mode symmetries in the measured Raman spectra at 300\,K (gray) by comparing with DFT simulated intensities (red). The modes are fitted using Lorentzian functions, and the fit is shown in black. (c-f) $T$-dependence of measured $\omega$ and $\Gamma$ for (c) A$_g$(1), (d) A$_g$(2), (e) A$_g$(5), and (f) A$_g$(7) modes. Error bars denote one standard deviation (s.d.). Eigenvectors of corresponding modes are shown next to each panel. Dashed linear lines are guide to the eye.
		}
		\label{fig:Phonons}
	\end{figure}

While the above-discussed momentum-averaged phonon dynamics revealed the material response in the entire Brillouin zone, the resolution was insufficient to decipher relatively small changes in $\omega$ and $\Gamma$. Hence, to identify and track individual phonon modes with significantly improved instrument resolution ($<$1\,cm$^{-1}$), we use $T-$dependent Raman scattering and simulations, and quantify $\Delta \omega$ and $\Delta \Gamma$ across $T_{\rm N}$. There are 42 Raman active [12$A_g$ + 12$B_{1g}$ + 9$B_{2g}$ + 9$B_{3g}$] modes at room-temperature in Bi$_2$Fe$_4$O$_9$ ($Pbam$ space group).~\cite{kroumova2003bilbao} We assign the measurable Raman peaks by comparing them with DFT simulated Raman intensity (see Fig.~\ref{fig:Phonons}(b))  and SM Table~S1). 
	
	\par Figure~\ref{fig:Phonons}(a) shows the color-map of Raman intensity from 15 to 800\,K. We do not find a visible discontinuity across $T_{\rm N}$. However, upon closer inspection, we find that $A_{g}$(1), $A_{g}$(2), $A_{g}$(5), and $A_{g}$(7) modes show a significant softening and anomalous broadening on heating (see $T-$dependence of $\omega$ and $\Gamma$ and mode eigenvectors in Fig.~\ref{fig:Phonons}(c-f)). Three key observations are to be noted here -- (i) distinct kinks are visible in both $\omega(T)$ and $\Gamma(T)$ at $T_{\rm N}$ and 1.8$T_{\rm N}$ (see orange and green dashed lines), (ii) a rapid drop in $\Gamma$ occurs at 1.8$T_{\rm N}$ instead of at $T_{\rm N}$ (see green dashed lines), and (iii) the low-energy Bi-dominated modes, i.e., $A_{g}$(1) and $A_{g}$(2), show a large $\Delta\omega/\omega$ of 6.9\% and 9.5\%, while the high-energy O-dominated vibrations, i.e., $A_{g}$(5) and $A_{g}$(7), change by 5.9\% and 3.7\% between 15 and 800\,K. The anomalous response in $\Gamma$ at 1.8$T_{\rm N}$ is also captured for some other modes, for example, $\Gamma$ of $B_{3g}$(1), $A_g$(3), and $B_{2g}$(4); however, it is not as pronounced (see $\Gamma$ vs.~$T$ in SM Fig.~S3). Note that only a few other modes, for example, $A_g$(4) and $B_{1g}$(2), show a distinct kink at $T_{\rm N}$ (see $\omega$ vs.~$T$ in SM Fig.~S3). From the above observations of lack of abrupt change in $\omega$ and $\Gamma$ at $T_{\rm N}$ within our instrument resolution, we can conclude relatively small measurable effects on phonons from the onset of long-range AFM order (magnetoelastic coupling and single- and double-ion anisotropy) and scattering of phonons from magnons or SRSF induced strain field above $T_{\rm N}$. Instead, the anomalous rise in $\Gamma$ between $T_{\rm N}$ and 1.8$T_{\rm N}$ suggests the possible role of other coupling mechanisms, which we explore below.
	
	\par
\begin{figure}[H]
		\begin{center}
			\includegraphics[trim=6cm 3.5cm 9.3cm 3.7cm, clip=true,width=0.8\textwidth]{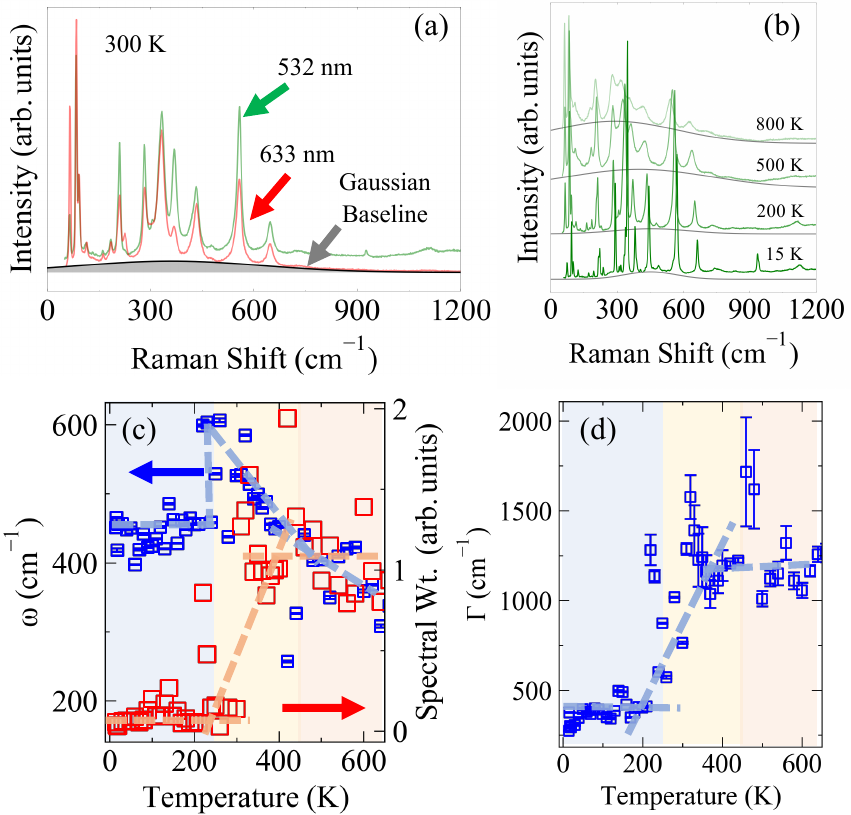}
		\end{center}
		\caption[Orbital fluctuations]{
			(a) Measurement of Raman intensity using two different pump laser excitations at 532 nm (green) and 633 nm (red). OFs are highlighted as Gaussian baseline (grey area) and remain at the same position for both pump lasers. A finite intensity above 900 cm$^{-1}$ corresponds to a PL peak identified in the extended energy-range scan (see SM Fig.~S4(b)). (b) $T$-dependence of Raman intensity highlighting the evolution of OFs. (c) The energy (blue squares) and spectral weight (red squares) of OFs as a function of temperature. (d) The linewidth of OFs as a function of temperature. The dashed lines are a guide to the eye in panels (c, d). Error bars, wherever shown, denote one s.d. or are smaller than marker size.
		}	
		\label{fig:Orbital_Fluctuations}
	\end{figure}

 On careful re-examination of our Raman data, we find a broad continuum centered near 400\,cm$^{-1}$ at 300\,K (see Fig.~\ref{fig:Orbital_Fluctuations}(a)). The broad continuum can also be seen in various probing configurations in previous polarized Raman scattering measurements,~\cite{iliev2010phonon}  in particular for the $z(yy)\overline{z}$ geometry, but has remained unidentified. On changing the excitation wavelength from 532 to 633\,nm, we find that the broad continuum's position does not shift (see Fig.~\ref{fig:Orbital_Fluctuations}(a)); hence, it can not be characterized as a photoluminescence (PL) signal. Moreover, the origin of a broad continuum can not be attributed to one-~\cite{kuroe2008magnetic} or two-~\cite{ulrich2006raman} magnons, as we know from our magnon measurements that one magnon excitations have narrow bandwidths and two magnon excitations originating from magnon bands near 20\,meV will have a maximum at 320\,cm$^{-1}$. Hence, considering the above observations and the absence of spin and charge degrees (band gap $\sim$2.1\,eV),~\cite{irshad2015first} we assign the broad continuum to OFs. The assignment is akin to the previous observation in LaTiO$_3$.~\cite{ulrich2006raman, ulrich2015spin}  On further extending the frequency range of our scans, second-order Raman peaks above 800\,cm$^{-1}$ are visible with 532\,nm excitation (see SM Fig.~S4(a)). This is due to strong Frank-Condon processes from Fe$^{3+}$ $d-d$ transition, as previously discussed in Ref.~\citenum{iliev2010phonon}. 
	
\par We now track the $T$-dependence of the broad continuum from 800 to 15\,K in Fig.~\ref{fig:Orbital_Fluctuations}(b). We find OFs are visible at 800\,K with large spectral weight and linewidth. On decresing temperature to 1.8$T_{\rm N}$, OFs energy stiffens from $\sim$320 to $\sim$420\,cm$^{-1}$, while the spectral weight and OFs linewidth nearly remain constant (see Fig.~\ref{fig:Orbital_Fluctuations}(c,d)). Between 1.8$T_{\rm N}$ and $T_{\rm N}$, we observe a significant renormalization of energy, spectral weight, and linewidth, which becomes nearly constant below $T_{\rm N}$. OFs linewidth is indicative of their intrinsic scattering and scattering with phonons. Hence, a significant drop below 1.8$T_{\rm N}$ suggests the suppression of scattering of OFs with the onset of SRSF. On stabilization of long-range AFM order below $T_{\rm N}$, OFs energy, linewidth, and spectral weight become constant, thus suggesting orbital confinement to spins below $T_{\rm N}$. The observed $T$-dependence of $\Gamma$ of phonons (Fig.~\ref{fig:Phonons}) and OFs (Fig.~\ref{fig:Orbital_Fluctuations}) is reminiscent of phonon scattering from OFs previously reported for LaTiO$_3$,~\cite{ulrich2015spin} CaBa(Co,Fe)$_4$O$_7$,~\cite{kocsis2023spin} and CrVO$_4$,~\cite{roy2024evidence} and occurs owing to the presence of KK-type interaction where the $d-$orbitals orientation confine to the Fe$^{3+}$ spins. 

\par Next, we focus on identifying the participating orbitals in OFs. Since the OF's suppression occurs on cooling near the onset of SRSF at 1.8$T_{\rm N}$ and long-range AFM order at $T_{\rm N}$, it is likely that OFs are from the partially filled Fe$^{3+}$ $d-$orbitals. Hence, if the OFs indeed originate from Fe-$d$ orbitals, we must also observe the renormalization of electronic states involving Fe-$d$ states. For possible observation of renormalization, we extend our frequency range and notice charge-transfer excitations (CTE) between Fe-$d$ and O-$p$ orbitals at $\sim$7500\,cm$^{-1}$ (0.9\,eV) and $\sim$4500\,cm$^{-1}$ (0.5\,eV) for 532\,nm (2.3\,eV) and 633\,nm (1.9\,eV) excitation wavelengths, respectively (see SM Fig.~S4(b)). Note that, since the observed signal changes with excitation wavelength (7500 to 4500\,cm$^{-1}$), it must correspond to PL at 1.4\,eV, which, with the aid of electronic structure simulations (see SM Fig.~S5), we can assign to the CTE between Fe-$d$ and O-$p$ orbitals. SM Fig.~S4(c) shows a strong energy renormalization of the CTE between 800 to 15\,K. In particular, the CTE remains nearly constant between 800\,K to 1.8\,$T_{\rm N}$, undergoes strong energy renormalization till $T_{\rm N}$, and again becomes constant between $T_{\rm N}$ and 15\,K. The same $T$-dependence of CTE energy and OFs confirms that OFs are indeed from partially filled Fe$^{3+}$ $d-$orbitals. Note that if we take a simple single-electron counting perspective, Fe$^{3+}$ with a 3$d^5$ configuration (exactly half filled) would give $S=2/5$, $L=0$ configuration by Hund's rule in the ground state and low-energy OFs would be suppressed as $m_L$ (or $L_z$, quantized projection of the angular momentum vector on the $z$-axis) can not change. Here $S$ is the spin, and $L$ is the orbital angular momentum. However, due to the complex bonding environment, there is a deviation from this simple picture, and electrons occupy both spin-up and spin-down states (see SM Fig.~S5), where OFs can exist. Here, we must emphasize that resonant inelastic X-ray scattering~\cite{ulrich2009momentum,lee2012two,ament2010theory} and resonant Raman scattering~\cite{kruger2004orbital,martin2004excitations,heyen1992resonant} can further shed light on orbital ordering and their fluctuations. In particular, Fe L-edge resonant inelastic X-ray scattering measurements in Bi$_2$Fe$_4$O$_9$ can reveal Fe$-d$ orbital ordering, their OFs owing to possible near degenerate states (similar to proposed OFs in Ref.~\citenum{kocsis2023spin} for Swedenborgites family), and the effect of polar lattice distortion on orbital ordering and fluctuations across $T_{\rm N}$. We will discuss the polar lattice distortion later in the text.

	\subsection{Role of Fe$_{\rm oct}$-O$_{\rm ap}$-Bi-O$_{\rm ap}$-Fe$_{\rm oct}$ Exchange on Anomalous Phonon Behavior}\label{sec:Bi_Role}
Beauvois et al.~\cite{beauvois2020dimer} discussed that the $J_{c}$, which connects the pentagonal planes and has the lowest magnitude among all $J^{\prime}$s, triggers the long-range magnetic transition to the AFM phase. Since $J_c$ plays a critical role in magnetic ordering, we focused on whether the observed anomalous behavior of $A_{g}$(1), $A_{g}$(2), $A_{g}$(5), and $A_{g}$(7) modes also possibly relates to it. 

\par $J_{c}$ has contribution from two exchange pathways -- Fe$_{\rm oct}$-O$_{\rm eq}$-Fe$_{\rm oct}$ and Fe$_{\rm oct}$-O$_{\rm ap}$-Bi-O$_{\rm ap}$-Fe$_{\rm oct}$ (see Fig.~\ref{fig:Crystal_Structure}(d) and discussion in Section~\ref{EIM}). If all the anomalous phonon modes strongly perturb any of the two exchange pathways and have a relatively weak effect on exchange interaction via Fe$_{\rm tet}$, we can ascertain that the anomalous phonons strongly couple to $J_c$. Both $A_{g}$(1) and $A_{g}$(2) modes have strong Bi displacement; thus they strongly perturb the Fe$_{\rm oct}$-O$_{\rm ap}$-Bi-O$_{\rm ap}$-Fe$_{\rm oct}$ pathway and consequently $J_c$. Since the effect of $A_{g}$(1) and $A_{g}$(2) modes is quite evident on $J_c$, we do not focus in detail on these two modes. Instead, we investigate the $A_{g}$(5), and $A_{g}$(7) modes that have primarlily O displacements. Note that the $A_g$(7) mode shows an in-phase breathing of O$_{\rm ap}$ atoms along the $a-$axis, and an out-of-phase stretching of O$_{\rm eq}$ atoms connecting the interlayer Fe$_{\rm oct}$ sites, while the $A_g$(5) mode primarily have a cross-plane motion of interlayer O$_{\rm eq}$ and displacement of Fe$_{\rm oct}$ and  O$_{\rm ap}$. SM Figures~S6 and~S7 show the $A_{g}$(5) and $A_{g}$(7) phonon perturbation's effect on electronic density of states (EDOS) of Fe$_{\rm tet}$, Fe$_{\rm oct}$, O$_{\rm ap}$, and O$_{\rm eq}$. As we can observe, for both the modes, Fe$_{\rm oct}$ and O$_{\rm ap}$ have strong modifications, while much weaker change occurs in EDOS of Fe$_{\rm tet}$ and O$_{\rm eq}$; thus suggesting the significant modification of Fe$_{\rm oct}$-O$_{\rm ap}$-Bi-O$_{\rm ap}$-Fe$_{\rm oct}$ exchange interaction as the exchange integral depends on integration of all energy states up to the Fermi level.~\cite{korotin2015calculation} Note that other exchange interactions, i.e., $J_{33}$, $J_{43}$, $J_{43}^{\prime}$, and $J_{44}$, which occurs either via Fe$_{\rm tet}$ or O$_{\rm eq}$, will have little change. From the above analysis, it is evident that all phonons showing pronounced anomalous behavior strongly perturb the Fe$_{\rm oct}$-O$_{\rm ap}$-Bi-O$_{\rm ap}$-Fe$_{\rm oct}$ exchange path and $J_c$.

\subsection{Determination of Spin-Phonon Coupling Constant}
As discussed in Section~\ref{APSS}, the $A_{g}$(1), $A_{g}$(2), $A_{g}$(5) and $A_{g}$(7) modes show a large $\Delta\omega/\omega$ of 6.9\%, 9.5\%, 5.9\% and 3.7\%, respectively, between 15 and 800\,K. These changes could originate from volume expansion, anharmonicity, or spin-phonon coupling. Within the quasiharmonic approximation, the volume expansion is related to the change in phonon energy as $\Delta \omega = -\gamma\omega\frac{\Delta V}{V}$, where $\gamma$ is the Gr\"uneisen parameter. We calculated the $\gamma$ using DFT simulations and values are listed in SM Table~S1. We find that for the modes considered here, $\Delta\omega$ is less than $\sim$4\,cm$^{-1}$ and can not account for observed $\Delta\omega$. Moreover, as evident from phonon broadening in the INS measurements and anharmonic $C_p$ (see Fig.~\ref{fig:Crystal_Structure}(b)), Bi$_2$Fe$_4$O$_9$ has large anharmonicity. In SM Fig.~S8, we fit the $\omega$-$T$ curve using expected variation from cubic anharmonicity~\cite{balkanski1983anharmonic} and calculate the deviation of anharmonic curves with the measured data at 15\,K, which can be attributed to spin-phonon coupling. Note that owing to the KK-type interaction discussed previously, the above approach to approximate the $\omega(T)$ using cubic anharmonicity is not adequate and results discussed below should be used only for qualitative interpretation. From the above difference at 15\,K and using the relationship $\Delta\omega_{sp} = -\frac{2.09}{m\omega}\frac{\partial^2J}{\partial{u^2}}\langle {\bm S}_i\cdot {\bm S}_j\rangle = -\lambda_{sp}S^2\phi(T)$,~\cite{kunwar2024raman, badola2024spin} we calculate the magnitude of $\lambda_{sp}$. Here, $m$ is the reduced mass, $\langle {\bm S}_i\cdot {\bm S}_j\rangle$ is spin-spin correlation factor, $S = \frac{5}{2}$ for Fe$^{3+}$, $u$ is phonon displacement, and $\phi(T)$ approximates the order parameter. In the mean-field approximation, $\phi(T)$ is zero above $T_{\rm N}$, but due to SRSF in Bi$_2$Fe$_4$O$_9$, it will be non-zero till 1.8$T_{\rm N}$. At 15\,K, we take $\phi$ to be 0.9. Using the above, we obtain $\lambda_{sp}$ to be 0.14, 0.23, 0.89, and 0.35\,cm$^{-1}$ for $A_g$(1), $A_g$(2), $A_g$(5), and $A_g$(7) modes, respectively. The $\lambda_{sp}$ values obtained for Bi$_2$Fe$_4$O$_9$ are comparable or smaller to previously reported spin-phonon coupled systems such as manganites, vanadates, and oxides, where values ranged from 0.5 to 5\,cm$^{-1}$.~\cite{granado1999magnetic,laverdiere2006spin,kunwar2024raman, badola2024spin} Hence, from the above qualitative analysis, we find Bi$_2$Fe$_4$O$_9$ to be a weakly spin-phonon coupled system. However, as shown earlier, Bi$_2$Fe$_4$O$_9$ harbors strong KK-type interaction, which led to a pronounced change in $\Gamma$ below 1.8$T_{\rm N}$.

	\subsection{Polar Distortion below $T_{\rm N}$}
	So far, we discussed the role of KK-type interaction and spin-phonon coupling, and realized various superexchange pathways in the frustrated Cairo lattice, Bi$_2$Fe$_4$O$_9$. Note that materials stabilizing in the Cairo lattice are inherently prone to magnetic frustrations due to an odd number of bonds.~\cite{beauvois2020dimer} Such frustrations are relieved via symmetry-lowering lattice distortions. Bi$_2$Fe$_4$O$_9$ closely resembles the $R$Mn$_2$O$_5$ family that also stabilizes in the Cairo lattice and undergoes a symmetry-lowering transition to a polar space group upon the onset of magnetic order.~\cite{cheong2007multiferroics,aguilar2006infrared,baledent2015evidence,chattopadhyay20173,garcia2007magnetically,kagomiya2003lattice,kim2011magnetic,zhao2014experimental} Such transitions manifest as an anomaly in dielectric measurements and spontaneous $\mathbf{P}$ in the FE phase.~\cite{kagomiya2003lattice,mihailova2005temperature,zhao2014experimental,chapon2004structural} Bi$_2$Fe$_4$O$_9$ holds an inversion symmetry in its orthorhombic phase at room temperature. On cooling, Singh et al.~\cite{singh2008substantial} reported spontaneous $\mathbf{P}$ and hysteresis near $T_{\rm N}$; however, the measured $\mathbf{P}\sim$2.5$\times$10$^{-3}$\,$\mu$Ccm$^{-2}$ at 0\,T is significantly smaller compared to $R$Mn$_2$O$_5$ (for example, $\mathbf{P} \sim $2\,$\mu$Ccm$^{-2}$ for DyMn$_2$O$_5$~\cite{zhao2014experimental}). Park \emph{et al}.~\cite{park2010effect} have also reported an anomaly in dielectric measurements at $T_{\rm N}$ along $b$ and $c-$axes. Thus, a measurable $\mathbf{P}$ is clearly indicative of a symmetry-lowering distortion that stabilizes the FE phase below $T_{\rm N}$. 
	
\par
\begin{figure}[H]
	 	\begin{center}
	 		\includegraphics[trim=2.01cm 4.0cm 11.75cm 3.5cm, clip=true, width=0.8\textwidth]{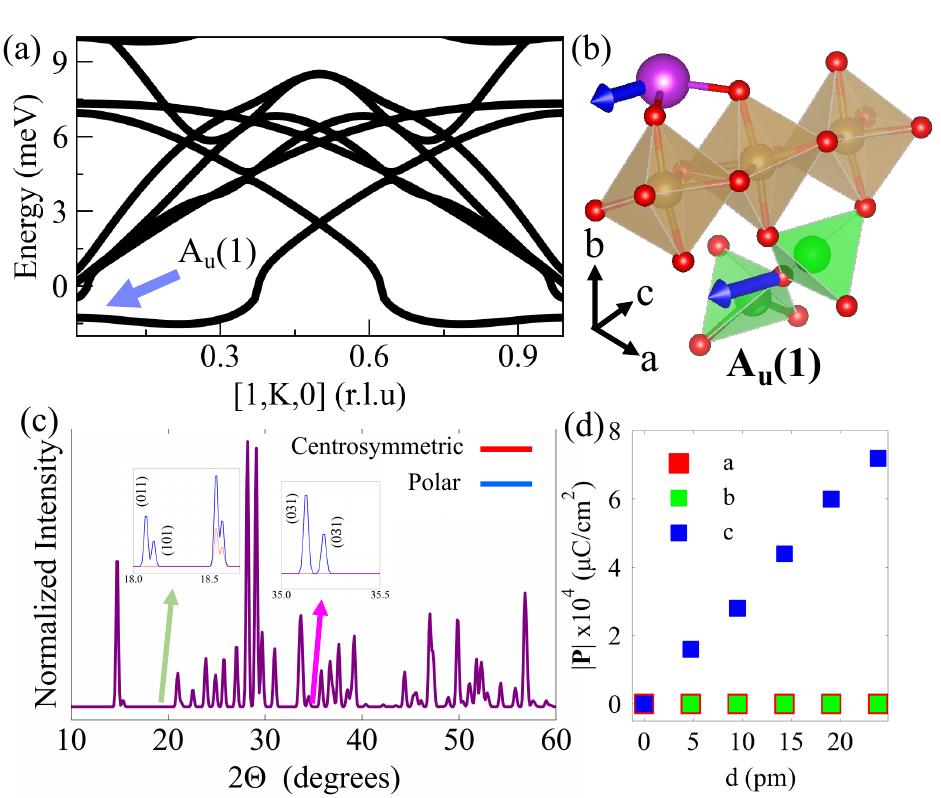}
	 	\end{center}
	 	\caption[Symmetry breaking from freezing of soft polar mode]{
	 		(a) Phonon dispersion of Bi$_2$Fe$_4$O$_9$ along [1, K, 0], showing the lowest unstable optical branch (indicated by the blue arrow). r.l.u.~refers to the reciprocal lattice units. (b) The eigenvector of the unstable phonon branch shown in panel (a) at the $\Gamma-$point. The mode is polar and has $A_u$(1) symmetry. (c) Simulated powder XRD of Bi$_2$Fe$_4$O$_9$ in the centrosymmetric (red) and distorted polar (blue) phase. For the polar phase, a small distortion average amplitude of $\sim$4.8\,pm was induced. Green and magenta arrows show the newly formed superlattice Bragg peaks in the polar phase. (d) Tracking the electronic component of the $\mathbf{P}$ upon perturbation with $A_u(1)$ mode along the three crystallographic axes.  A measurable change in $\mathbf{P}$ along the $c-$axis is visible, while along the $a$ and $b$ axes, $\mathbf{P}$ is nearly zero.
	 	}	
	 	\label{fig:Symmetry_Break}
	 \end{figure}

 To investigate a possible lattice distortion, we revisit our INS measurements. A comparison between Fig.~\ref{fig:Spin_Wave}(c) and (d) highlights the appearance of a measurable intensity captured close to $E\sim$3\,meV at 5\,K. Integrated $\chi^{\prime\prime}(E)$ (see Fig.~\ref{fig:Phonon_DOS}(c)) confirms the characteristic peak near 3\,meV. The peak becomes broad on heating but is visible in our data at least till 130\,K (not shown). Using longitudinal polarization analysis, the same feature captured in earlier INS measurements~\cite{le2021experimental} was attributed to a have a non-magnetic origin. The 3\,meV mode, which is not visible above $T_{\rm N}$, possibly arises from a lattice instability below $T_{\rm N}$. We obtain the phonon dispersion from our $ab$ $initio$ simulations to investigate the instability further. The simulated dispersion captures a weak instability in the lowest optic branch along [0,0,0] to [$\frac{1}{2},\frac{1}{2},\frac{1}{2}$] in large part of the Brillouin zone (see Fig.~\ref{fig:Symmetry_Break}(a)). On careful analysis of the mode symmetry of the unstable branch, we find that at the zone-center, the mode has $A_u$(1) symmetry and is polar in nature. Figure~\ref{fig:Symmetry_Break}(b) shows the corresponding eigenvector of the mode showing a dominant Bi and O displacement along the $c-$axis. Upon freezing the $A_u$(1) mode, we obtain a new structure with a polar space group. Symmetry lowering polar lattice distortion at $\Gamma-$point manifests as superlattice Bragg peaks below $T_{\rm N}$ that were symmetry forbidden above $T_{\rm N}$. We simulate the diffraction pattern in Bi$_2$Fe$_4$O$_9$ for both the phases (see Fig.~\ref{fig:Symmetry_Break}(c)) and identify the superlattice Bragg peaks. The calculated intensities are extremely small (proportional to the square of the $A_u$(1) distortion amplitude) and are yet to be experimentally confirmed. Previous neutron diffraction measurements, possibly due to their weak intensity, also did not report the superlattice Bragg peaks.~\cite{ressouche2009magnetic} A high-flux and high-resolution single-crystal diffraction study is necessary to confirm or deny the stabilization of the distorted polar phase. 
	
 \par Next, to confirm the experimentally reported magnitude of $\mathbf{P}$, we perturb the room-temperature orthorhombic phase with $A_u$(1) distortion. Note that the distortion involves a Bi displacement along the $c$-axis that displaces the Bi-$6s$ lone pair along the same direction to induce $\mathbf{P}$. We calculate the electronic contribution to the induced $\mathbf{P}$ using the Berry phase method implemented in VASP. Here, we integrate with respect to the center of the $c$-axis along the Bi coordinates. The $\mathbf{P}$ is shown in Figure~\ref{fig:Symmetry_Break}(d). We observe a linear change of the induced $\mathbf{P}$ upon $A_u$(1) mode perturbation, and the magnitude is comparable to the measured value~\cite{singh2008substantial} for maximum perturbation considered here. Hence, the above analysis suggests that Bi atom off-centering induced by polar distortion is critical in stabilizing the FE phase. 
 
\par In previous Section~\ref{sec:Bi_Role}), we observed that due to KK-type interaction the phonons perturbing Fe$_{\rm oct}$-O$_{\rm ap}$-Bi-O$_{\rm ap}$-Fe$_{\rm oct}$ superexchange pathway exhibited anomalies in linewidth below 1.8\,$T_{\rm N}$. The unstable polar $A_u$(1) distortion also perturbs the Fe$_{\rm oct}$-O$_{\rm ap}$-Bi-O$_{\rm ap}$-Fe$_{\rm oct}$ pathway; hence, it is possible that polar distortion originates from KK-type interaction. If that is the case, similar to other discussed modes, the $A_u$(1) distortion will also show an anomalous response below 1.8$T_{\rm N}$ and ${\mathbf P}$ arising from $A_u$(1) distortion will be visible between $T_{\rm N}$ and 1.8$T_{\rm N}$. Indeed, the $\mathbf{P}$ hysteresis reported in Ref.~\citenum{singh2008substantial} was at 250\,K (slightly above $T_{\rm N}$) and showed strong coupling to the external magnetic field. However, due to the small value of $\mathbf{P}$, their measurements of $\mathbf{P}$ hysteresis at room temperature were limited by leakage current. New sensitive $\mathbf{P}$ measurements and theoretical modeling of KK-type interaction and polar mode instability can possibly elucidate this point further.
 		 
	\subsection{Thermal Transport}
	Magnetic semiconductors and insulators, for example, YMnO$_{3}$~\cite{sharma2004thermal} and CrN~\cite{stockem2018anomalous} show a pronounced change in $\kappa$ across $T_{\rm N}$ owing to strong phonon scattering from strain induced by SRSF. In Section~\ref{APSS}, we discussed a large increase in $\Gamma$ between $T_{\rm N}$ and 1.8$T_{\rm N}$ owing to KK-type interaction; although we had excluded the role of SRSF due to the absence of sudden discontinuity at $T_{\rm N}$. It remains to be investigated whether KK-type interaction-induced change in $\Gamma$ for selected modes could also lead to discontinuity or a change of slope in $\kappa$ at $T_{\rm N}$.

\par
\begin{figure}[H]
		\begin{center}
			\includegraphics[trim=1.31cm 4.55cm 5.15cm 2.6cm, clip=true, width=0.7\textwidth]{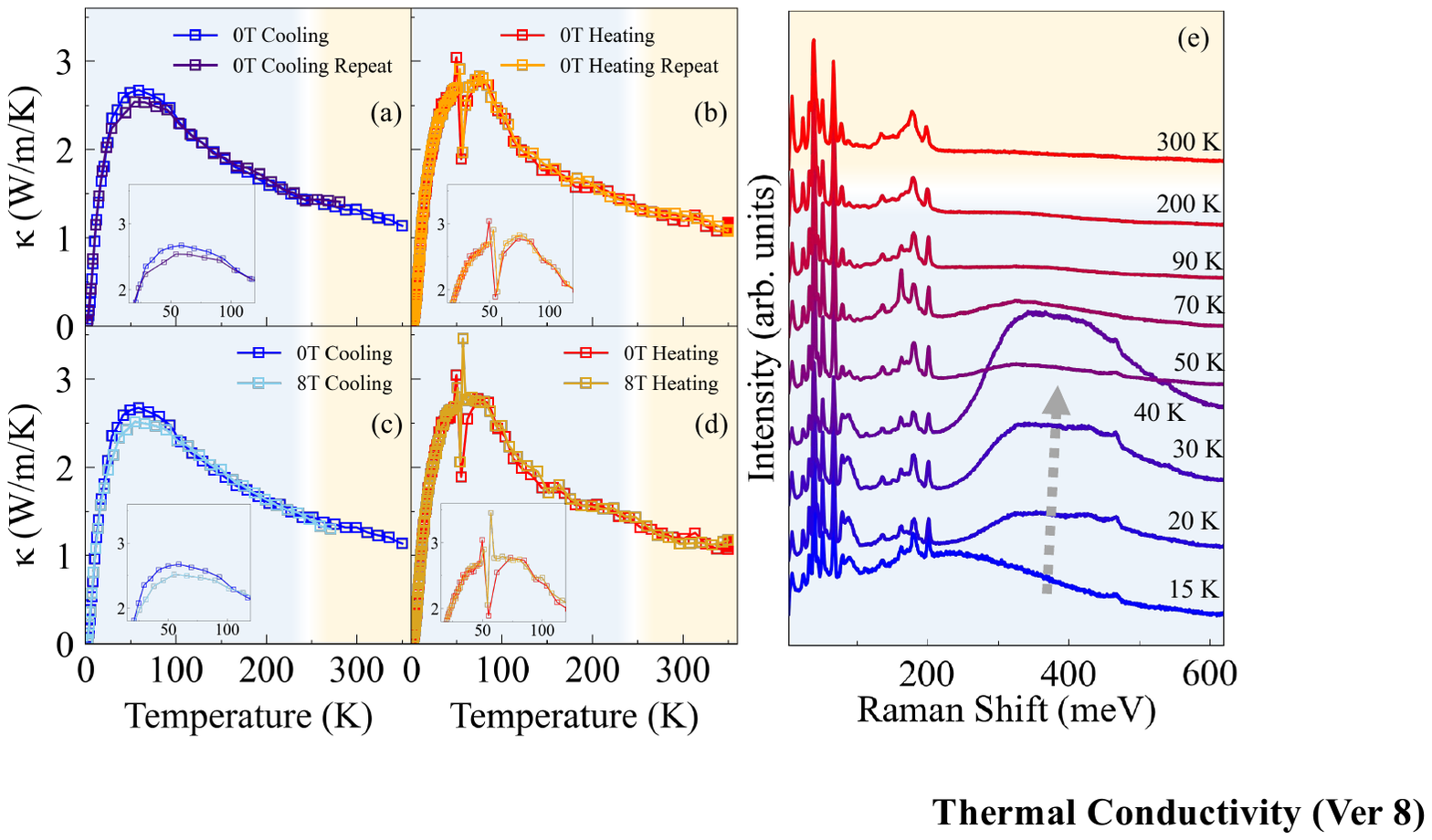}
		\end{center}
		\caption{
			Thermal conductivity ($\kappa$) of Bi$_2$Fe$_4$O$_9$ measured on both cooling (panel a) and heating (panel b) cycles. The repeat measurements are from a different run on the same sample. A visible anomaly at $\sim$57\,K on the heating cycle is further zoomed in the inset of panel (b). (c,d) Comparison of $\kappa$ measurements at $B$ = 0 and 8\,T on cooling and heating cycles. Zoomed inset in panel (d) highlights that the anomaly is independent of $B$. Error bars are smaller than marker size. (e) Extended energy-range scans of Raman intensity highlighting the presence of peak near 400\,meV between 15 and 50\,K, which diminishes in intensity on further heating. 
		}
		\label{fig:Kappa}
	\end{figure}

 Figures~\ref{fig:Kappa}(a) and~\ref{fig:Kappa}(b) show the $\kappa$ measurements spanning 2-350\,K in zero-field on cooling and heating cycles, respectively. We repeated the measurements in a different run to confirm the reproducibility of the data. Two features are apparent from the data -- (i) an anomaly at $\sim$57\,K on the heating cycle, which is absent in the cooling cycle, and (ii) a hint of slope change near $T_{\rm N}$ (but within the error bars that are of approximately same size as markers). The absence of distinct change at $T_{\rm N}$ rules out any measurable effect of KK-type interaction on $\kappa$. However, the anomaly at $\sim$57\,K is quite intriguing since neither the previous magnetic susceptibility, diffraction, and magnon measurements on Bi$_2$Fe$_4$O$_9$,~\cite{ressouche2009magnetic, beauvois2020dimer, le2021experimental} nor our $C_p$, Raman scattering, and elastic and inelastic neutron scattering showed this feature. This anomaly persists at least up to 8\,T and further rules out any field-induced spin reorientation or phase transition (Figure~\ref{fig:Kappa}(c,d)). We revisited our Raman measurements of OFs and PL peaks. Interestingly, Fig.~\ref{fig:Kappa}(e) shows a visible broad continuum peak near $\sim$400\,meV between 15 and 50\,K with a large spectral weight that diminishes above 50\,K. Given a large electronic gap of $\sim$2.1\,eV and absence of phonons and magnons at such high energy, we speculate that the observed $\sim$400\,meV continuum possibly arises from an orbital reorientation of Fe$^{3+}$ $d-$states. We further rule out any structural change as we do not observe the appearance or disappearance of phonon peaks or find any measurable change in phonon peak energy or intensity in Raman measurements. The observed feature in both $\kappa$ and Raman measurements on the heating cycle requires detailed theoretical modeling, which is beyond the scope of the present study. 	
	
	\section{Summary}
\par In summary, through a combination of INS and Raman measurements complemented with DFT simulations, we affirm that Bi$_2$Fe$_4$O$_9$ manifests a strong KK-type interaction, i.e., a coupled spin-orbital phase below 1.8$T_{\rm N}$. The KK-type interaction is facilitated by orbital degrees originating from partially filled Fe$^{3+}$ $d-$orbitals. We confirm the role of orbital degrees by measuring the low-energy orbital excitations and their subsequent $T-$dependence. In the coupled spin-orbital phase, the phonons interact with the orbital degrees via perturbation of the Fe$_{\rm oct}$-O$_{\rm ap}$-Bi-O$_{\rm ap}$-Fe$_{\rm oct}$ superexchange pathway leading to an order of magnitude change in observed phonon linewidths. Notably, a small change in phonon energy arising from long-range AFM order below $T_{\rm N}$ highlights a weak spin-phonon coupling. Moreover, we identify the origin of previously reported $\mathbf{P}$ hysteresis in Bi$_2$Fe$_4$O$_9$. We find the symmetry-lowering polar $A_u$(1) distortion exhibiting a $c-$axis motion of Bi atoms displaces the Bi$-6s$ loan pair and induces $\mathbf{P}$. It is likely that the $A_u$(1) distortion is triggered by KK-type interaction. Moreover, the KK-type interactions do not lead to any significant slope change in the temperature dependence of $\kappa$. Instead, the $T$-dependent $\kappa$ uncovers an anomalous spike $\sim$57\,K appearing only in the heating cycle. Our Raman measurements also show an anomalous response of broad $\sim$400\,meV continuum in the same temperature range. Further measurements and theoretical modeling are required to understand the anomaly. Our results highlight the role of KK-type interaction as a promising avenue for realizing strong ME coupling.

	\section*{Acknowledgements}
	The authors acknowledge Dr. Vivek Dwij and Dr. Vasant Sathe for extending their support with preliminary measurements of Raman spectroscopy at CSR-UGC Indore. A.P.R. acknowledges the financial support from IRCC-IITB. D.B. thanks the financial support from the Science \& Engineering Research Board (SERB) under Project No. CRG/2022/001317 and MoE/STARS under Project No. MoE/STARS-1/345. The simulations were performed in the SPACETIME-II super-computing facility at IITB. The neutron scattering experiments were carried out at the MARI beamline at the ISIS facility, supported by beamtime allocation RB2190107.  Raman spectroscopy measurements were performed at the Laser Raman Imaging System (central facility) at IIT Bombay.
	
\newpage	

\begin{thebibliography}{63}%
\bibitem [{\citenamefont {Wannier}(1950)}]{wannier1950antiferromagnetism}%
  \BibitemOpen
  \bibfield  {author} {\bibinfo {author} {\bibfnamefont {G.}~\bibnamefont
  {Wannier}},\ }\bibfield  {title} {\bibinfo {title} {Antiferromagnetism. the
  triangular ising net},\ }\href@noop {} {\bibfield  {journal} {\bibinfo
  {journal} {Physical Review}\ }\textbf {\bibinfo {volume} {79}},\ \bibinfo
  {pages} {357} (\bibinfo {year} {1950})}\BibitemShut {NoStop}%
\bibitem [{\citenamefont {Newnham}\ \emph {et~al.}(1978)\citenamefont
  {Newnham}, \citenamefont {Kramer}, \citenamefont {Schulze},\ and\
  \citenamefont {Cross}}]{newnham1978magnetoferroelectricity}%
  \BibitemOpen
  \bibfield  {author} {\bibinfo {author} {\bibfnamefont {R.~E.}\ \bibnamefont
  {Newnham}}, \bibinfo {author} {\bibfnamefont {J.}~\bibnamefont {Kramer}},
  \bibinfo {author} {\bibfnamefont {W.}~\bibnamefont {Schulze}},\ and\ \bibinfo
  {author} {\bibfnamefont {L.}~\bibnamefont {Cross}},\ }\bibfield  {title}
  {\bibinfo {title} {Magnetoferroelectricity in {Cr$_2$BeO$_4$}},\ }\href@noop
  {} {\bibfield  {journal} {\bibinfo  {journal} {Journal of Applied Physics}\
  }\textbf {\bibinfo {volume} {49}},\ \bibinfo {pages} {6088} (\bibinfo {year}
  {1978})}\BibitemShut {NoStop}%
\bibitem [{\citenamefont {Fiebig}\ \emph {et~al.}(2016)\citenamefont {Fiebig},
  \citenamefont {Lottermoser}, \citenamefont {Meier},\ and\ \citenamefont
  {Trassin}}]{fiebig2016evolution}%
  \BibitemOpen
  \bibfield  {author} {\bibinfo {author} {\bibfnamefont {M.}~\bibnamefont
  {Fiebig}}, \bibinfo {author} {\bibfnamefont {T.}~\bibnamefont {Lottermoser}},
  \bibinfo {author} {\bibfnamefont {D.}~\bibnamefont {Meier}},\ and\ \bibinfo
  {author} {\bibfnamefont {M.}~\bibnamefont {Trassin}},\ }\bibfield  {title}
  {\bibinfo {title} {The evolution of multiferroics},\ }\href@noop {}
  {\bibfield  {journal} {\bibinfo  {journal} {Nature Reviews Materials}\
  }\textbf {\bibinfo {volume} {1}},\ \bibinfo {pages} {1} (\bibinfo {year}
  {2016})}\BibitemShut {NoStop}%
\bibitem [{\citenamefont {Arima}\ \emph {et~al.}(2006)\citenamefont {Arima},
  \citenamefont {Tokunaga}, \citenamefont {Goto}, \citenamefont {Kimura},
  \citenamefont {Noda},\ and\ \citenamefont {Tokura}}]{arima2006collinear}%
  \BibitemOpen
  \bibfield  {author} {\bibinfo {author} {\bibfnamefont {T.}~\bibnamefont
  {Arima}}, \bibinfo {author} {\bibfnamefont {A.}~\bibnamefont {Tokunaga}},
  \bibinfo {author} {\bibfnamefont {T.}~\bibnamefont {Goto}}, \bibinfo {author}
  {\bibfnamefont {H.}~\bibnamefont {Kimura}}, \bibinfo {author} {\bibfnamefont
  {Y.}~\bibnamefont {Noda}},\ and\ \bibinfo {author} {\bibfnamefont
  {Y.}~\bibnamefont {Tokura}},\ }\bibfield  {title} {\bibinfo {title}
  {Collinear to spiral spin transformation without changing the modulation
  wavelength upon ferroelectric transition in {Tb$_{1- x}$Dy$_x$MnO$_3$}},\
  }\href@noop {} {\bibfield  {journal} {\bibinfo  {journal} {Physical Review
  Letters}\ }\textbf {\bibinfo {volume} {96}},\ \bibinfo {pages} {097202}
  (\bibinfo {year} {2006})}\BibitemShut {NoStop}%
\bibitem [{\citenamefont {Katsura}\ \emph {et~al.}(2005)\citenamefont
  {Katsura}, \citenamefont {Nagaosa},\ and\ \citenamefont
  {Balatsky}}]{katsura2005spin}%
  \BibitemOpen
  \bibfield  {author} {\bibinfo {author} {\bibfnamefont {H.}~\bibnamefont
  {Katsura}}, \bibinfo {author} {\bibfnamefont {N.}~\bibnamefont {Nagaosa}},\
  and\ \bibinfo {author} {\bibfnamefont {A.~V.}\ \bibnamefont {Balatsky}},\
  }\bibfield  {title} {\bibinfo {title} {Spin current and magnetoelectric
  effect in noncollinear magnets},\ }\href@noop {} {\bibfield  {journal}
  {\bibinfo  {journal} {Physical Review Letters}\ }\textbf {\bibinfo {volume}
  {95}},\ \bibinfo {pages} {057205} (\bibinfo {year} {2005})}\BibitemShut
  {NoStop}%
\bibitem [{\citenamefont {Mostovoy}(2006)}]{mostovoy2006ferroelectricity}%
  \BibitemOpen
  \bibfield  {author} {\bibinfo {author} {\bibfnamefont {M.}~\bibnamefont
  {Mostovoy}},\ }\bibfield  {title} {\bibinfo {title} {Ferroelectricity in
  spiral magnets},\ }\href@noop {} {\bibfield  {journal} {\bibinfo  {journal}
  {Physical Review Letters}\ }\textbf {\bibinfo {volume} {96}},\ \bibinfo
  {pages} {067601} (\bibinfo {year} {2006})}\BibitemShut {NoStop}%
\bibitem [{\citenamefont {Kocsis}\ \emph {et~al.}(2023)\citenamefont {Kocsis},
  \citenamefont {Tokunaga}, \citenamefont {R{\~o}{\~o}m}, \citenamefont
  {Nagel}, \citenamefont {Fujioka}, \citenamefont {Taguchi}, \citenamefont
  {Tokura},\ and\ \citenamefont {Bord{\'a}cs}}]{kocsis2023spin}%
  \BibitemOpen
  \bibfield  {author} {\bibinfo {author} {\bibfnamefont {V.}~\bibnamefont
  {Kocsis}}, \bibinfo {author} {\bibfnamefont {Y.}~\bibnamefont {Tokunaga}},
  \bibinfo {author} {\bibfnamefont {T.}~\bibnamefont {R{\~o}{\~o}m}}, \bibinfo
  {author} {\bibfnamefont {U.}~\bibnamefont {Nagel}}, \bibinfo {author}
  {\bibfnamefont {J.}~\bibnamefont {Fujioka}}, \bibinfo {author} {\bibfnamefont
  {Y.}~\bibnamefont {Taguchi}}, \bibinfo {author} {\bibfnamefont
  {Y.}~\bibnamefont {Tokura}},\ and\ \bibinfo {author} {\bibfnamefont
  {S.}~\bibnamefont {Bord{\'a}cs}},\ }\bibfield  {title} {\bibinfo {title}
  {Spin-lattice and magnetoelectric couplings enhanced by orbital degrees of
  freedom in polar multiferroic semiconductors},\ }\href@noop {} {\bibfield
  {journal} {\bibinfo  {journal} {Physical Review Letters}\ }\textbf {\bibinfo
  {volume} {130}},\ \bibinfo {pages} {036801} (\bibinfo {year}
  {2023})}\BibitemShut {NoStop}%
\bibitem [{\citenamefont {Johnson}\ \emph {et~al.}(2012)\citenamefont
  {Johnson}, \citenamefont {Chapon}, \citenamefont {Khalyavin}, \citenamefont
  {Manuel}, \citenamefont {Radaelli},\ and\ \citenamefont
  {Martin}}]{johnson2012giant}%
  \BibitemOpen
  \bibfield  {author} {\bibinfo {author} {\bibfnamefont {R.}~\bibnamefont
  {Johnson}}, \bibinfo {author} {\bibfnamefont {L.}~\bibnamefont {Chapon}},
  \bibinfo {author} {\bibfnamefont {D.}~\bibnamefont {Khalyavin}}, \bibinfo
  {author} {\bibfnamefont {P.}~\bibnamefont {Manuel}}, \bibinfo {author}
  {\bibfnamefont {P.}~\bibnamefont {Radaelli}},\ and\ \bibinfo {author}
  {\bibfnamefont {C.}~\bibnamefont {Martin}},\ }\bibfield  {title} {\bibinfo
  {title} {Giant improper ferroelectricity in the ferroaxial magnet
  {CaMn$_7$O$_{12}$}},\ }\href@noop {} {\bibfield  {journal} {\bibinfo
  {journal} {Physical Review Letters}\ }\textbf {\bibinfo {volume} {108}},\
  \bibinfo {pages} {067201} (\bibinfo {year} {2012})}\BibitemShut {NoStop}%
\bibitem [{\citenamefont {Kugel}\ and\ \citenamefont
  {Khomskii}(1973)}]{kugel1973crystal}%
  \BibitemOpen
  \bibfield  {author} {\bibinfo {author} {\bibfnamefont {K.}~\bibnamefont
  {Kugel}}\ and\ \bibinfo {author} {\bibfnamefont {D.}~\bibnamefont
  {Khomskii}},\ }\bibfield  {title} {\bibinfo {title} {Crystal-structure and
  magnetic properties of substances with orbital degeneracy},\ }\href@noop {}
  {\bibfield  {journal} {\bibinfo  {journal} {Zh. Eksp. Teor. Fiz}\ }\textbf
  {\bibinfo {volume} {64}},\ \bibinfo {pages} {1429} (\bibinfo {year}
  {1973})}\BibitemShut {NoStop}%
\bibitem [{\citenamefont {Khomskii}\ and\ \citenamefont
  {Mostovoy}(2003)}]{khomskii2003orbital}%
  \BibitemOpen
  \bibfield  {author} {\bibinfo {author} {\bibfnamefont {D.}~\bibnamefont
  {Khomskii}}\ and\ \bibinfo {author} {\bibfnamefont {M.}~\bibnamefont
  {Mostovoy}},\ }\bibfield  {title} {\bibinfo {title} {Orbital ordering and
  frustrations},\ }\href@noop {} {\bibfield  {journal} {\bibinfo  {journal}
  {Journal of Physics A: Mathematical and General}\ }\textbf {\bibinfo {volume}
  {36}},\ \bibinfo {pages} {9197} (\bibinfo {year} {2003})}\BibitemShut
  {NoStop}%
\bibitem [{\citenamefont {Roy}\ \emph {et~al.}(2024)\citenamefont {Roy},
  \citenamefont {Jayakrishnan}, \citenamefont {Dwij}, \citenamefont
  {Khandelwal}, \citenamefont {Chattopadhyay}, \citenamefont {Sathe},
  \citenamefont {Mittal}, \citenamefont {Sastry}, \citenamefont {Achary},
  \citenamefont {Tyagi} \emph {et~al.}}]{roy2024evidence}%
  \BibitemOpen
  \bibfield  {author} {\bibinfo {author} {\bibfnamefont {A.~P.}\ \bibnamefont
  {Roy}}, \bibinfo {author} {\bibfnamefont {S.}~\bibnamefont {Jayakrishnan}},
  \bibinfo {author} {\bibfnamefont {V.}~\bibnamefont {Dwij}}, \bibinfo {author}
  {\bibfnamefont {A.}~\bibnamefont {Khandelwal}}, \bibinfo {author}
  {\bibfnamefont {M.}~\bibnamefont {Chattopadhyay}}, \bibinfo {author}
  {\bibfnamefont {V.}~\bibnamefont {Sathe}}, \bibinfo {author} {\bibfnamefont
  {R.}~\bibnamefont {Mittal}}, \bibinfo {author} {\bibfnamefont
  {P.}~\bibnamefont {Sastry}}, \bibinfo {author} {\bibfnamefont {S.~N.}\
  \bibnamefont {Achary}}, \bibinfo {author} {\bibfnamefont {A.~K.}\
  \bibnamefont {Tyagi}}, \emph {et~al.},\ }\bibfield  {title} {\bibinfo {title}
  {Evidence of strong orbital-selective spin-orbital-phonon coupling in
  {CrVO$_4$}},\ }\href@noop {} {\bibfield  {journal} {\bibinfo  {journal}
  {Physical Review Letters}\ }\textbf {\bibinfo {volume} {132}},\ \bibinfo
  {pages} {026701} (\bibinfo {year} {2024})}\BibitemShut {NoStop}%
\bibitem [{\citenamefont {Disa}\ \emph {et~al.}(2023)\citenamefont {Disa},
  \citenamefont {Curtis}, \citenamefont {Fechner}, \citenamefont {Liu},
  \citenamefont {von Hoegen}, \citenamefont {F{\"o}rst}, \citenamefont {Nova},
  \citenamefont {Narang}, \citenamefont {Maljuk}, \citenamefont {Boris} \emph
  {et~al.}}]{disa2023photo}%
  \BibitemOpen
  \bibfield  {author} {\bibinfo {author} {\bibfnamefont {A.}~\bibnamefont
  {Disa}}, \bibinfo {author} {\bibfnamefont {J.}~\bibnamefont {Curtis}},
  \bibinfo {author} {\bibfnamefont {M.}~\bibnamefont {Fechner}}, \bibinfo
  {author} {\bibfnamefont {A.}~\bibnamefont {Liu}}, \bibinfo {author}
  {\bibfnamefont {A.}~\bibnamefont {von Hoegen}}, \bibinfo {author}
  {\bibfnamefont {M.}~\bibnamefont {F{\"o}rst}}, \bibinfo {author}
  {\bibfnamefont {T.}~\bibnamefont {Nova}}, \bibinfo {author} {\bibfnamefont
  {P.}~\bibnamefont {Narang}}, \bibinfo {author} {\bibfnamefont
  {A.}~\bibnamefont {Maljuk}}, \bibinfo {author} {\bibfnamefont
  {A.}~\bibnamefont {Boris}}, \emph {et~al.},\ }\bibfield  {title} {\bibinfo
  {title} {Photo-induced high-temperature ferromagnetism in {YTiO$_3$}},\
  }\href@noop {} {\bibfield  {journal} {\bibinfo  {journal} {Nature}\ }\textbf
  {\bibinfo {volume} {617}},\ \bibinfo {pages} {73} (\bibinfo {year}
  {2023})}\BibitemShut {NoStop}%
\bibitem [{\citenamefont {Lee}\ \emph {et~al.}(2012)\citenamefont {Lee},
  \citenamefont {Yuan}, \citenamefont {Lal}, \citenamefont {Joe}, \citenamefont
  {Gan}, \citenamefont {Smadici}, \citenamefont {Finkelstein}, \citenamefont
  {Feng}, \citenamefont {Rusydi}, \citenamefont {Goldbart} \emph
  {et~al.}}]{lee2012two}%
  \BibitemOpen
  \bibfield  {author} {\bibinfo {author} {\bibfnamefont {J.~C.}\ \bibnamefont
  {Lee}}, \bibinfo {author} {\bibfnamefont {S.}~\bibnamefont {Yuan}}, \bibinfo
  {author} {\bibfnamefont {S.}~\bibnamefont {Lal}}, \bibinfo {author}
  {\bibfnamefont {Y.~I.}\ \bibnamefont {Joe}}, \bibinfo {author} {\bibfnamefont
  {Y.}~\bibnamefont {Gan}}, \bibinfo {author} {\bibfnamefont {S.}~\bibnamefont
  {Smadici}}, \bibinfo {author} {\bibfnamefont {K.}~\bibnamefont
  {Finkelstein}}, \bibinfo {author} {\bibfnamefont {Y.}~\bibnamefont {Feng}},
  \bibinfo {author} {\bibfnamefont {A.}~\bibnamefont {Rusydi}}, \bibinfo
  {author} {\bibfnamefont {P.~M.}\ \bibnamefont {Goldbart}}, \emph {et~al.},\
  }\bibfield  {title} {\bibinfo {title} {Two-stage orbital order and dynamical
  spin frustration in {KCuF$_3$}},\ }\href@noop {} {\bibfield  {journal}
  {\bibinfo  {journal} {Nature Physics}\ }\textbf {\bibinfo {volume} {8}},\
  \bibinfo {pages} {63} (\bibinfo {year} {2012})}\BibitemShut {NoStop}%
\bibitem [{\citenamefont {Keimer}\ \emph {et~al.}(2000)\citenamefont {Keimer},
  \citenamefont {Casa}, \citenamefont {Ivanov}, \citenamefont {Lynn},
  \citenamefont {Zimmermann}, \citenamefont {Hill}, \citenamefont {Gibbs},
  \citenamefont {Taguchi},\ and\ \citenamefont {Tokura}}]{keimer2000spin}%
  \BibitemOpen
  \bibfield  {author} {\bibinfo {author} {\bibfnamefont {B.}~\bibnamefont
  {Keimer}}, \bibinfo {author} {\bibfnamefont {D.}~\bibnamefont {Casa}},
  \bibinfo {author} {\bibfnamefont {A.}~\bibnamefont {Ivanov}}, \bibinfo
  {author} {\bibfnamefont {J.}~\bibnamefont {Lynn}}, \bibinfo {author}
  {\bibfnamefont {M.~v.}\ \bibnamefont {Zimmermann}}, \bibinfo {author}
  {\bibfnamefont {J.}~\bibnamefont {Hill}}, \bibinfo {author} {\bibfnamefont
  {D.}~\bibnamefont {Gibbs}}, \bibinfo {author} {\bibfnamefont
  {Y.}~\bibnamefont {Taguchi}},\ and\ \bibinfo {author} {\bibfnamefont
  {Y.}~\bibnamefont {Tokura}},\ }\bibfield  {title} {\bibinfo {title} {Spin
  dynamics and orbital state in {LaTiO$_3$}},\ }\href@noop {} {\bibfield
  {journal} {\bibinfo  {journal} {Physical Review Letters}\ }\textbf {\bibinfo
  {volume} {85}},\ \bibinfo {pages} {3946} (\bibinfo {year}
  {2000})}\BibitemShut {NoStop}%
\bibitem [{\citenamefont {Ulrich}\ \emph {et~al.}(2006)\citenamefont {Ulrich},
  \citenamefont {G{\"o}ssling}, \citenamefont {Gr{\"u}ninger}, \citenamefont
  {Guennou}, \citenamefont {Roth}, \citenamefont {Cwik}, \citenamefont
  {Lorenz}, \citenamefont {Khaliullin},\ and\ \citenamefont
  {Keimer}}]{ulrich2006raman}%
  \BibitemOpen
  \bibfield  {author} {\bibinfo {author} {\bibfnamefont {C.}~\bibnamefont
  {Ulrich}}, \bibinfo {author} {\bibfnamefont {A.}~\bibnamefont
  {G{\"o}ssling}}, \bibinfo {author} {\bibfnamefont {M.}~\bibnamefont
  {Gr{\"u}ninger}}, \bibinfo {author} {\bibfnamefont {M.}~\bibnamefont
  {Guennou}}, \bibinfo {author} {\bibfnamefont {H.}~\bibnamefont {Roth}},
  \bibinfo {author} {\bibfnamefont {M.}~\bibnamefont {Cwik}}, \bibinfo {author}
  {\bibfnamefont {T.}~\bibnamefont {Lorenz}}, \bibinfo {author} {\bibfnamefont
  {G.}~\bibnamefont {Khaliullin}},\ and\ \bibinfo {author} {\bibfnamefont
  {B.}~\bibnamefont {Keimer}},\ }\bibfield  {title} {\bibinfo {title} {Raman
  scattering in the mott insulators {LaTiO$_3$} and {YTiO$_3$:} evidence for
  orbital excitations},\ }\href@noop {} {\bibfield  {journal} {\bibinfo
  {journal} {Physical Review Letters}\ }\textbf {\bibinfo {volume} {97}},\
  \bibinfo {pages} {157401} (\bibinfo {year} {2006})}\BibitemShut {NoStop}%
\bibitem [{\citenamefont {Ulrich}\ \emph {et~al.}(2009)\citenamefont {Ulrich},
  \citenamefont {Ament}, \citenamefont {Ghiringhelli}, \citenamefont
  {Braicovich}, \citenamefont {Moretti~Sala}, \citenamefont {Pezzotta},
  \citenamefont {Schmitt}, \citenamefont {Khaliullin}, \citenamefont {Van
  Den~Brink}, \citenamefont {Roth} \emph {et~al.}}]{ulrich2009momentum}%
  \BibitemOpen
  \bibfield  {author} {\bibinfo {author} {\bibfnamefont {C.}~\bibnamefont
  {Ulrich}}, \bibinfo {author} {\bibfnamefont {L.}~\bibnamefont {Ament}},
  \bibinfo {author} {\bibfnamefont {G.}~\bibnamefont {Ghiringhelli}}, \bibinfo
  {author} {\bibfnamefont {L.}~\bibnamefont {Braicovich}}, \bibinfo {author}
  {\bibfnamefont {M.}~\bibnamefont {Moretti~Sala}}, \bibinfo {author}
  {\bibfnamefont {N.}~\bibnamefont {Pezzotta}}, \bibinfo {author}
  {\bibfnamefont {T.}~\bibnamefont {Schmitt}}, \bibinfo {author} {\bibfnamefont
  {G.}~\bibnamefont {Khaliullin}}, \bibinfo {author} {\bibfnamefont
  {J.}~\bibnamefont {Van Den~Brink}}, \bibinfo {author} {\bibfnamefont
  {H.}~\bibnamefont {Roth}}, \emph {et~al.},\ }\bibfield  {title} {\bibinfo
  {title} {Momentum dependence of orbital excitations in mott-insulating
  titanates},\ }\href@noop {} {\bibfield  {journal} {\bibinfo  {journal}
  {Physical Review Letters}\ }\textbf {\bibinfo {volume} {103}},\ \bibinfo
  {pages} {107205} (\bibinfo {year} {2009})}\BibitemShut {NoStop}%
\bibitem [{\citenamefont {Ulrich}\ \emph {et~al.}(2015)\citenamefont {Ulrich},
  \citenamefont {Khaliullin}, \citenamefont {Guennou}, \citenamefont {Roth},
  \citenamefont {Lorenz},\ and\ \citenamefont {Keimer}}]{ulrich2015spin}%
  \BibitemOpen
  \bibfield  {author} {\bibinfo {author} {\bibfnamefont {C.}~\bibnamefont
  {Ulrich}}, \bibinfo {author} {\bibfnamefont {G.}~\bibnamefont {Khaliullin}},
  \bibinfo {author} {\bibfnamefont {M.}~\bibnamefont {Guennou}}, \bibinfo
  {author} {\bibfnamefont {H.}~\bibnamefont {Roth}}, \bibinfo {author}
  {\bibfnamefont {T.}~\bibnamefont {Lorenz}},\ and\ \bibinfo {author}
  {\bibfnamefont {B.}~\bibnamefont {Keimer}},\ }\bibfield  {title} {\bibinfo
  {title} {Spin-orbital excitation continuum and anomalous electron-phonon
  interaction in the mott insulator {LaTiO$_3$}},\ }\href@noop {} {\bibfield
  {journal} {\bibinfo  {journal} {Physical review letters}\ }\textbf {\bibinfo
  {volume} {115}},\ \bibinfo {pages} {156403} (\bibinfo {year}
  {2015})}\BibitemShut {NoStop}%
\bibitem [{\citenamefont {Miyasaka}\ \emph {et~al.}(2005)\citenamefont
  {Miyasaka}, \citenamefont {Onoda}, \citenamefont {Okimoto}, \citenamefont
  {Fujioka}, \citenamefont {Iwama}, \citenamefont {Nagaosa},\ and\
  \citenamefont {Tokura}}]{miyasaka2005one}%
  \BibitemOpen
  \bibfield  {author} {\bibinfo {author} {\bibfnamefont {S.}~\bibnamefont
  {Miyasaka}}, \bibinfo {author} {\bibfnamefont {S.}~\bibnamefont {Onoda}},
  \bibinfo {author} {\bibfnamefont {Y.}~\bibnamefont {Okimoto}}, \bibinfo
  {author} {\bibfnamefont {J.}~\bibnamefont {Fujioka}}, \bibinfo {author}
  {\bibfnamefont {M.}~\bibnamefont {Iwama}}, \bibinfo {author} {\bibfnamefont
  {N.}~\bibnamefont {Nagaosa}},\ and\ \bibinfo {author} {\bibfnamefont
  {Y.}~\bibnamefont {Tokura}},\ }\bibfield  {title} {\bibinfo {title}
  {One-dimensional orbital excitations in vanadium oxides},\ }\href@noop {}
  {\bibfield  {journal} {\bibinfo  {journal} {Physical Review Letters}\
  }\textbf {\bibinfo {volume} {94}},\ \bibinfo {pages} {076405} (\bibinfo
  {year} {2005})}\BibitemShut {NoStop}%
\bibitem [{\citenamefont {Sugai}\ and\ \citenamefont
  {Hirota}(2006)}]{sugai2006orbital}%
  \BibitemOpen
  \bibfield  {author} {\bibinfo {author} {\bibfnamefont {S.}~\bibnamefont
  {Sugai}}\ and\ \bibinfo {author} {\bibfnamefont {K.}~\bibnamefont {Hirota}},\
  }\bibfield  {title} {\bibinfo {title} {Orbital waves in {YVO$_3$} studied by
  {Raman} scattering},\ }\href@noop {} {\bibfield  {journal} {\bibinfo
  {journal} {Physical Review B}\ }\textbf {\bibinfo {volume} {73}},\ \bibinfo
  {pages} {020409} (\bibinfo {year} {2006})}\BibitemShut {NoStop}%
\bibitem [{\citenamefont {Ressouche}\ \emph {et~al.}(2009)\citenamefont
  {Ressouche}, \citenamefont {Simonet}, \citenamefont {Canals}, \citenamefont
  {Gospodinov},\ and\ \citenamefont {Skumryev}}]{ressouche2009magnetic}%
  \BibitemOpen
  \bibfield  {author} {\bibinfo {author} {\bibfnamefont {E.}~\bibnamefont
  {Ressouche}}, \bibinfo {author} {\bibfnamefont {V.}~\bibnamefont {Simonet}},
  \bibinfo {author} {\bibfnamefont {B.}~\bibnamefont {Canals}}, \bibinfo
  {author} {\bibfnamefont {M.}~\bibnamefont {Gospodinov}},\ and\ \bibinfo
  {author} {\bibfnamefont {V.}~\bibnamefont {Skumryev}},\ }\bibfield  {title}
  {\bibinfo {title} {Magnetic frustration in an iron-based {Cairo} pentagonal
  lattice},\ }\href@noop {} {\bibfield  {journal} {\bibinfo  {journal}
  {Physical Review Letters}\ }\textbf {\bibinfo {volume} {103}},\ \bibinfo
  {pages} {267204} (\bibinfo {year} {2009})}\BibitemShut {NoStop}%
\bibitem [{\citenamefont {Singh}\ \emph {et~al.}(2008)\citenamefont {Singh},
  \citenamefont {Kaushik}, \citenamefont {Kumar}, \citenamefont {Mishra},
  \citenamefont {Venimadhav}, \citenamefont {Siruguri},\ and\ \citenamefont
  {Patnaik}}]{singh2008substantial}%
  \BibitemOpen
  \bibfield  {author} {\bibinfo {author} {\bibfnamefont {A.}~\bibnamefont
  {Singh}}, \bibinfo {author} {\bibfnamefont {S.}~\bibnamefont {Kaushik}},
  \bibinfo {author} {\bibfnamefont {B.}~\bibnamefont {Kumar}}, \bibinfo
  {author} {\bibfnamefont {P.}~\bibnamefont {Mishra}}, \bibinfo {author}
  {\bibfnamefont {A.}~\bibnamefont {Venimadhav}}, \bibinfo {author}
  {\bibfnamefont {V.}~\bibnamefont {Siruguri}},\ and\ \bibinfo {author}
  {\bibfnamefont {S.}~\bibnamefont {Patnaik}},\ }\bibfield  {title} {\bibinfo
  {title} {Substantial magnetoelectric coupling near room temperature in
  {Bi$_2$Fe$_4$O$_9$}},\ }\href@noop {} {\bibfield  {journal} {\bibinfo
  {journal} {Applied Physics Letters}\ }\textbf {\bibinfo {volume} {92}}
  (\bibinfo {year} {2008})}\BibitemShut {NoStop}%
\bibitem [{\citenamefont {Toby}\ and\ \citenamefont
  {Von~Dreele}(2013)}]{toby2013gsas}%
  \BibitemOpen
  \bibfield  {author} {\bibinfo {author} {\bibfnamefont {B.~H.}\ \bibnamefont
  {Toby}}\ and\ \bibinfo {author} {\bibfnamefont {R.~B.}\ \bibnamefont
  {Von~Dreele}},\ }\bibfield  {title} {\bibinfo {title} {{GSAS-II}: the genesis
  of a modern open-source all purpose crystallography software package},\
  }\href@noop {} {\bibfield  {journal} {\bibinfo  {journal} {Journal of Applied
  Crystallography}\ }\textbf {\bibinfo {volume} {46}},\ \bibinfo {pages} {544}
  (\bibinfo {year} {2013})}\BibitemShut {NoStop}%
\bibitem [{SM_()}]{SM_ref_Bi2Fe4O9}%
  \BibitemOpen
  \href@noop {} {\bibinfo  {journal} {See Supplemental Material
  {\color{blue}[url]} for Figures S1-S8, and Table S1}\ }\BibitemShut {NoStop}%
\bibitem [{\citenamefont {Shamir}\ \emph {et~al.}(1978)\citenamefont {Shamir},
  \citenamefont {Gurewitz},\ and\ \citenamefont {Shaked}}]{shamir1978magnetic}%
  \BibitemOpen
\bibfield  {journal} {  }\bibfield  {author} {\bibinfo {author} {\bibfnamefont
  {N.}~\bibnamefont {Shamir}}, \bibinfo {author} {\bibfnamefont
  {E.}~\bibnamefont {Gurewitz}},\ and\ \bibinfo {author} {\bibfnamefont
  {H.}~\bibnamefont {Shaked}},\ }\bibfield  {title} {\bibinfo {title} {The
  magnetic structure of {Bi$_2$Fe$_4$O$_9$}--analysis of neutron diffraction
  measurements},\ }\href@noop {} {\bibfield  {journal} {\bibinfo  {journal}
  {Acta Crystallographica Section A: Crystal Physics, Diffraction, Theoretical
  and General Crystallography}\ }\textbf {\bibinfo {volume} {34}},\ \bibinfo
  {pages} {662} (\bibinfo {year} {1978})}\BibitemShut {NoStop}%
\bibitem [{\citenamefont {Arnold}\ \emph {et~al.}(2014)\citenamefont {Arnold},
  \citenamefont {Bilheux}, \citenamefont {Borreguero}, \citenamefont {Buts},
  \citenamefont {Campbell}, \citenamefont {Chapon}, \citenamefont {Doucet},
  \citenamefont {Draper}, \citenamefont {Leal}, \citenamefont {Gigg} \emph
  {et~al.}}]{arnold2014mantid}%
  \BibitemOpen
  \bibfield  {author} {\bibinfo {author} {\bibfnamefont {O.}~\bibnamefont
  {Arnold}}, \bibinfo {author} {\bibfnamefont {J.-C.}\ \bibnamefont {Bilheux}},
  \bibinfo {author} {\bibfnamefont {J.}~\bibnamefont {Borreguero}}, \bibinfo
  {author} {\bibfnamefont {A.}~\bibnamefont {Buts}}, \bibinfo {author}
  {\bibfnamefont {S.~I.}\ \bibnamefont {Campbell}}, \bibinfo {author}
  {\bibfnamefont {L.}~\bibnamefont {Chapon}}, \bibinfo {author} {\bibfnamefont
  {M.}~\bibnamefont {Doucet}}, \bibinfo {author} {\bibfnamefont
  {N.}~\bibnamefont {Draper}}, \bibinfo {author} {\bibfnamefont {R.~F.}\
  \bibnamefont {Leal}}, \bibinfo {author} {\bibfnamefont {M.}~\bibnamefont
  {Gigg}}, \emph {et~al.},\ }\bibfield  {title} {\bibinfo {title}
  {Mantid—data analysis and visualization package for neutron scattering and
  {$\mu$-SR} experiments},\ }\href@noop {} {\bibfield  {journal} {\bibinfo
  {journal} {Nuclear Instruments and Methods in Physics Research Section A:
  Accelerators, Spectrometers, Detectors and Associated Equipment}\ }\textbf
  {\bibinfo {volume} {764}},\ \bibinfo {pages} {156} (\bibinfo {year}
  {2014})}\BibitemShut {NoStop}%
\bibitem [{\citenamefont {Bansal}\ \emph {et~al.}(2015)\citenamefont {Bansal},
  \citenamefont {Li}, \citenamefont {Said}, \citenamefont {Abernathy},
  \citenamefont {Yan},\ and\ \citenamefont {Delaire}}]{bansal2015electron}%
  \BibitemOpen
  \bibfield  {author} {\bibinfo {author} {\bibfnamefont {D.}~\bibnamefont
  {Bansal}}, \bibinfo {author} {\bibfnamefont {C.~W.}\ \bibnamefont {Li}},
  \bibinfo {author} {\bibfnamefont {A.~H.}\ \bibnamefont {Said}}, \bibinfo
  {author} {\bibfnamefont {D.~L.}\ \bibnamefont {Abernathy}}, \bibinfo {author}
  {\bibfnamefont {J.}~\bibnamefont {Yan}},\ and\ \bibinfo {author}
  {\bibfnamefont {O.}~\bibnamefont {Delaire}},\ }\bibfield  {title} {\bibinfo
  {title} {Electron-phonon coupling and thermal transport in the thermoelectric
  compound {$\mathrm{Mo_{3}Sb_{7-x}Te_x}$}},\ }\href@noop {} {\bibfield
  {journal} {\bibinfo  {journal} {Physical Review B}\ }\textbf {\bibinfo
  {volume} {92}},\ \bibinfo {pages} {214301} (\bibinfo {year}
  {2015})}\BibitemShut {NoStop}%
\bibitem [{\citenamefont {Kresse}\ and\ \citenamefont
  {Hafner}(1993)}]{kresse1993ab}%
  \BibitemOpen
  \bibfield  {author} {\bibinfo {author} {\bibfnamefont {G.}~\bibnamefont
  {Kresse}}\ and\ \bibinfo {author} {\bibfnamefont {J.}~\bibnamefont
  {Hafner}},\ }\bibfield  {title} {\bibinfo {title} {Abinitio molecular
  dynamics for open-shell transition metals},\ }\href@noop {} {\bibfield
  {journal} {\bibinfo  {journal} {Physical Review B}\ }\textbf {\bibinfo
  {volume} {48}},\ \bibinfo {pages} {13115} (\bibinfo {year}
  {1993})}\BibitemShut {NoStop}%
\bibitem [{\citenamefont {Kresse}\ and\ \citenamefont
  {Furthm{\"u}ller}(1996{\natexlab{a}})}]{kresse1996efficient}%
  \BibitemOpen
  \bibfield  {author} {\bibinfo {author} {\bibfnamefont {G.}~\bibnamefont
  {Kresse}}\ and\ \bibinfo {author} {\bibfnamefont {J.}~\bibnamefont
  {Furthm{\"u}ller}},\ }\bibfield  {title} {\bibinfo {title} {Efficient
  iterative schemes for ab-initio total-energy calculations using a plane-wave
  basis set},\ }\href@noop {} {\bibfield  {journal} {\bibinfo  {journal}
  {Physical Review B}\ }\textbf {\bibinfo {volume} {54}},\ \bibinfo {pages}
  {11169} (\bibinfo {year} {1996}{\natexlab{a}})}\BibitemShut {NoStop}%
\bibitem [{\citenamefont {Kresse}\ and\ \citenamefont
  {Furthm{\"u}ller}(1996{\natexlab{b}})}]{kresse1996efficiency}%
  \BibitemOpen
  \bibfield  {author} {\bibinfo {author} {\bibfnamefont {G.}~\bibnamefont
  {Kresse}}\ and\ \bibinfo {author} {\bibfnamefont {J.}~\bibnamefont
  {Furthm{\"u}ller}},\ }\bibfield  {title} {\bibinfo {title} {Efficiency of
  abinitio total energy calculations for metals and semiconductors using a
  plane-wave basis set},\ }\href@noop {} {\bibfield  {journal} {\bibinfo
  {journal} {Computational Materials Science}\ }\textbf {\bibinfo {volume}
  {6}},\ \bibinfo {pages} {15} (\bibinfo {year}
  {1996}{\natexlab{b}})}\BibitemShut {NoStop}%
\bibitem [{\citenamefont {Csonka}\ \emph {et~al.}(2009)\citenamefont {Csonka},
  \citenamefont {Perdew}, \citenamefont {Ruzsinszky}, \citenamefont
  {Philipsen}, \citenamefont {Leb{\`e}gue}, \citenamefont {Paier},
  \citenamefont {Vydrov},\ and\ \citenamefont
  {{\'A}ngy{\'a}n}}]{csonka2009assessing}%
  \BibitemOpen
  \bibfield  {author} {\bibinfo {author} {\bibfnamefont {G.~I.}\ \bibnamefont
  {Csonka}}, \bibinfo {author} {\bibfnamefont {J.~P.}\ \bibnamefont {Perdew}},
  \bibinfo {author} {\bibfnamefont {A.}~\bibnamefont {Ruzsinszky}}, \bibinfo
  {author} {\bibfnamefont {P.~H.}\ \bibnamefont {Philipsen}}, \bibinfo {author}
  {\bibfnamefont {S.}~\bibnamefont {Leb{\`e}gue}}, \bibinfo {author}
  {\bibfnamefont {J.}~\bibnamefont {Paier}}, \bibinfo {author} {\bibfnamefont
  {O.~A.}\ \bibnamefont {Vydrov}},\ and\ \bibinfo {author} {\bibfnamefont
  {J.~G.}\ \bibnamefont {{\'A}ngy{\'a}n}},\ }\bibfield  {title} {\bibinfo
  {title} {Assessing the performance of recent density functionals for bulk
  solids},\ }\href@noop {} {\bibfield  {journal} {\bibinfo  {journal} {Physical
  Review B}\ }\textbf {\bibinfo {volume} {79}},\ \bibinfo {pages} {155107}
  (\bibinfo {year} {2009})}\BibitemShut {NoStop}%
\bibitem [{\citenamefont {Dudarev}\ \emph {et~al.}(1998)\citenamefont
  {Dudarev}, \citenamefont {Botton}, \citenamefont {Savrasov}, \citenamefont
  {Humphreys},\ and\ \citenamefont {Sutton}}]{dudarev1998electron}%
  \BibitemOpen
  \bibfield  {author} {\bibinfo {author} {\bibfnamefont {S.~L.}\ \bibnamefont
  {Dudarev}}, \bibinfo {author} {\bibfnamefont {G.~A.}\ \bibnamefont {Botton}},
  \bibinfo {author} {\bibfnamefont {S.~Y.}\ \bibnamefont {Savrasov}}, \bibinfo
  {author} {\bibfnamefont {C.}~\bibnamefont {Humphreys}},\ and\ \bibinfo
  {author} {\bibfnamefont {A.~P.}\ \bibnamefont {Sutton}},\ }\bibfield  {title}
  {\bibinfo {title} {Electron-energy-loss spectra and the structural stability
  of nickel oxide: {An} {LSDA+U} study},\ }\href@noop {} {\bibfield  {journal}
  {\bibinfo  {journal} {Physical Review B}\ }\textbf {\bibinfo {volume} {57}},\
  \bibinfo {pages} {1505} (\bibinfo {year} {1998})}\BibitemShut {NoStop}%
\bibitem [{\citenamefont {Irshad}\ \emph {et~al.}(2015)\citenamefont {Irshad},
  \citenamefont {Shah}, \citenamefont {Rafiq},\ and\ \citenamefont
  {Hasan}}]{irshad2015first}%
  \BibitemOpen
  \bibfield  {author} {\bibinfo {author} {\bibfnamefont {Z.}~\bibnamefont
  {Irshad}}, \bibinfo {author} {\bibfnamefont {S.}~\bibnamefont {Shah}},
  \bibinfo {author} {\bibfnamefont {M.}~\bibnamefont {Rafiq}},\ and\ \bibinfo
  {author} {\bibfnamefont {M.}~\bibnamefont {Hasan}},\ }\bibfield  {title}
  {\bibinfo {title} {First principles study of structural, electronic and
  magnetic properties of ferromagnetic {Bi$_2$Fe$_4$O$_9$}},\ }\href@noop {}
  {\bibfield  {journal} {\bibinfo  {journal} {Journal of Alloys and Compounds}\
  }\textbf {\bibinfo {volume} {624}},\ \bibinfo {pages} {131} (\bibinfo {year}
  {2015})}\BibitemShut {NoStop}%
\bibitem [{\citenamefont {Togo}\ and\ \citenamefont
  {Tanaka}(2015)}]{togo2015first}%
  \BibitemOpen
  \bibfield  {author} {\bibinfo {author} {\bibfnamefont {A.}~\bibnamefont
  {Togo}}\ and\ \bibinfo {author} {\bibfnamefont {I.}~\bibnamefont {Tanaka}},\
  }\bibfield  {title} {\bibinfo {title} {First principles phonon calculations
  in materials science},\ }\href@noop {} {\bibfield  {journal} {\bibinfo
  {journal} {Scripta Materialia}\ }\textbf {\bibinfo {volume} {108}},\ \bibinfo
  {pages} {1} (\bibinfo {year} {2015})}\BibitemShut {NoStop}%
\bibitem [{\citenamefont {Skelton}\ \emph {et~al.}(2017)\citenamefont
  {Skelton}, \citenamefont {Burton}, \citenamefont {Jackson}, \citenamefont
  {Oba}, \citenamefont {Parker},\ and\ \citenamefont
  {Walsh}}]{skelton2017lattice}%
  \BibitemOpen
  \bibfield  {author} {\bibinfo {author} {\bibfnamefont {J.~M.}\ \bibnamefont
  {Skelton}}, \bibinfo {author} {\bibfnamefont {L.~A.}\ \bibnamefont {Burton}},
  \bibinfo {author} {\bibfnamefont {A.~J.}\ \bibnamefont {Jackson}}, \bibinfo
  {author} {\bibfnamefont {F.}~\bibnamefont {Oba}}, \bibinfo {author}
  {\bibfnamefont {S.~C.}\ \bibnamefont {Parker}},\ and\ \bibinfo {author}
  {\bibfnamefont {A.}~\bibnamefont {Walsh}},\ }\bibfield  {title} {\bibinfo
  {title} {Lattice dynamics of the tin sulphides {SnS$_2$}, {SnS} and
  {Sn$_2$S$_3$}: vibrational spectra and thermal transport},\ }\href@noop {}
  {\bibfield  {journal} {\bibinfo  {journal} {Physical Chemistry Chemical
  Physics}\ }\textbf {\bibinfo {volume} {19}},\ \bibinfo {pages} {12452}
  (\bibinfo {year} {2017})}\BibitemShut {NoStop}%
\bibitem [{\citenamefont {Toth}\ and\ \citenamefont
  {Lake}(2015)}]{toth2015linear}%
  \BibitemOpen
  \bibfield  {author} {\bibinfo {author} {\bibfnamefont {S.}~\bibnamefont
  {Toth}}\ and\ \bibinfo {author} {\bibfnamefont {B.}~\bibnamefont {Lake}},\
  }\bibfield  {title} {\bibinfo {title} {Linear spin wave theory for {single-Q}
  incommensurate magnetic structures},\ }\href@noop {} {\bibfield  {journal}
  {\bibinfo  {journal} {Journal of Physics: Condensed Matter}\ }\textbf
  {\bibinfo {volume} {27}},\ \bibinfo {pages} {166002} (\bibinfo {year}
  {2015})}\BibitemShut {NoStop}%
\bibitem [{\citenamefont {Le}\ \emph {et~al.}(2021)\citenamefont {Le},
  \citenamefont {Wheeler}, \citenamefont {Jeong}, \citenamefont {Kumar},
  \citenamefont {Lee}, \citenamefont {Lee}, \citenamefont {Oh}, \citenamefont
  {Jo}, \citenamefont {Kondo}, \citenamefont {Kindo} \emph
  {et~al.}}]{le2021experimental}%
  \BibitemOpen
  \bibfield  {author} {\bibinfo {author} {\bibfnamefont {M.~D.}\ \bibnamefont
  {Le}}, \bibinfo {author} {\bibfnamefont {E.~M.}\ \bibnamefont {Wheeler}},
  \bibinfo {author} {\bibfnamefont {J.}~\bibnamefont {Jeong}}, \bibinfo
  {author} {\bibfnamefont {K.~R.}\ \bibnamefont {Kumar}}, \bibinfo {author}
  {\bibfnamefont {S.}~\bibnamefont {Lee}}, \bibinfo {author} {\bibfnamefont
  {C.-H.}\ \bibnamefont {Lee}}, \bibinfo {author} {\bibfnamefont {M.~J.}\
  \bibnamefont {Oh}}, \bibinfo {author} {\bibfnamefont {Y.-J.}\ \bibnamefont
  {Jo}}, \bibinfo {author} {\bibfnamefont {A.}~\bibnamefont {Kondo}}, \bibinfo
  {author} {\bibfnamefont {K.}~\bibnamefont {Kindo}}, \emph {et~al.},\
  }\bibfield  {title} {\bibinfo {title} {Experimental determination of the
  magnetic interactions of frustrated cairo pentagon lattice materials},\
  }\href@noop {} {\bibfield  {journal} {\bibinfo  {journal} {Physical Review
  B}\ }\textbf {\bibinfo {volume} {103}},\ \bibinfo {pages} {104423} (\bibinfo
  {year} {2021})}\BibitemShut {NoStop}%
\bibitem [{\citenamefont {Beauvois}\ \emph {et~al.}(2020)\citenamefont
  {Beauvois}, \citenamefont {Simonet}, \citenamefont {Petit}, \citenamefont
  {Robert}, \citenamefont {Bourdarot}, \citenamefont {Gospodinov},
  \citenamefont {Mukhin}, \citenamefont {Ballou}, \citenamefont {Skumryev},\
  and\ \citenamefont {Ressouche}}]{beauvois2020dimer}%
  \BibitemOpen
  \bibfield  {author} {\bibinfo {author} {\bibfnamefont {K.}~\bibnamefont
  {Beauvois}}, \bibinfo {author} {\bibfnamefont {V.}~\bibnamefont {Simonet}},
  \bibinfo {author} {\bibfnamefont {S.}~\bibnamefont {Petit}}, \bibinfo
  {author} {\bibfnamefont {J.}~\bibnamefont {Robert}}, \bibinfo {author}
  {\bibfnamefont {F.}~\bibnamefont {Bourdarot}}, \bibinfo {author}
  {\bibfnamefont {M.}~\bibnamefont {Gospodinov}}, \bibinfo {author}
  {\bibfnamefont {A.}~\bibnamefont {Mukhin}}, \bibinfo {author} {\bibfnamefont
  {R.}~\bibnamefont {Ballou}}, \bibinfo {author} {\bibfnamefont
  {V.}~\bibnamefont {Skumryev}},\ and\ \bibinfo {author} {\bibfnamefont
  {E.}~\bibnamefont {Ressouche}},\ }\bibfield  {title} {\bibinfo {title} {Dimer
  physics in the frustrated cairo pentagonal antiferromagnet
  {Bi$_2$Fe$_4$O$_9$}},\ }\href@noop {} {\bibfield  {journal} {\bibinfo
  {journal} {Physical Review Letters}\ }\textbf {\bibinfo {volume} {124}},\
  \bibinfo {pages} {127202} (\bibinfo {year} {2020})}\BibitemShut {NoStop}%
\bibitem [{\citenamefont {Geertsma}\ and\ \citenamefont
  {Khomskii}(1996)}]{geertsma1996influence}%
  \BibitemOpen
  \bibfield  {author} {\bibinfo {author} {\bibfnamefont {W.}~\bibnamefont
  {Geertsma}}\ and\ \bibinfo {author} {\bibfnamefont {D.}~\bibnamefont
  {Khomskii}},\ }\bibfield  {title} {\bibinfo {title} {Influence of side groups
  on {90$^{\circ}$} superexchange: A modification of the
  {Goodenough-Kanamori-Anderson} rules},\ }\href@noop {} {\bibfield  {journal}
  {\bibinfo  {journal} {Physical Review B}\ }\textbf {\bibinfo {volume} {54}},\
  \bibinfo {pages} {3011} (\bibinfo {year} {1996})}\BibitemShut {NoStop}%
\bibitem [{\citenamefont {Pchelkina}\ and\ \citenamefont
  {Streltsov}(2013)}]{pchelkina2013ab}%
  \BibitemOpen
  \bibfield  {author} {\bibinfo {author} {\bibfnamefont {Z.}~\bibnamefont
  {Pchelkina}}\ and\ \bibinfo {author} {\bibfnamefont {S.}~\bibnamefont
  {Streltsov}},\ }\bibfield  {title} {\bibinfo {title} {Ab initio investigation
  of the exchange interactions in {Bi$_2$Fe$_4$O$_9$:} the {Cairo} pentagonal
  lattice compound},\ }\href@noop {} {\bibfield  {journal} {\bibinfo  {journal}
  {Physical Review B}\ }\textbf {\bibinfo {volume} {88}},\ \bibinfo {pages}
  {054424} (\bibinfo {year} {2013})}\BibitemShut {NoStop}%
\bibitem [{\citenamefont {Gippius}\ \emph {et~al.}(1998)\citenamefont
  {Gippius}, \citenamefont {VasiL'Ev}, \citenamefont {Petrakovskii},
  \citenamefont {Zalessky}, \citenamefont {Hoffmann}, \citenamefont
  {L{\"u}ders}, \citenamefont {Dhalenne},\ and\ \citenamefont
  {Revcolevschi}}]{gippius1998observation}%
  \BibitemOpen
  \bibfield  {author} {\bibinfo {author} {\bibfnamefont {A.}~\bibnamefont
  {Gippius}}, \bibinfo {author} {\bibfnamefont {A.}~\bibnamefont {VasiL'Ev}},
  \bibinfo {author} {\bibfnamefont {G.}~\bibnamefont {Petrakovskii}}, \bibinfo
  {author} {\bibfnamefont {A.}~\bibnamefont {Zalessky}}, \bibinfo {author}
  {\bibfnamefont {W.}~\bibnamefont {Hoffmann}}, \bibinfo {author}
  {\bibfnamefont {K.}~\bibnamefont {L{\"u}ders}}, \bibinfo {author}
  {\bibfnamefont {G.}~\bibnamefont {Dhalenne}},\ and\ \bibinfo {author}
  {\bibfnamefont {A.}~\bibnamefont {Revcolevschi}},\ }\bibfield  {title}
  {\bibinfo {title} {Observation of $^{63, 65}${Cu} and $^{209}${Bi} nuclear
  resonance in antiferromagnetic {Bi$_2$CuO$_4$}},\ }\href@noop {} {\bibfield
  {journal} {\bibinfo  {journal} {Journal of Magnetism and Magnetic Materials}\
  }\textbf {\bibinfo {volume} {184}},\ \bibinfo {pages} {358} (\bibinfo {year}
  {1998})}\BibitemShut {NoStop}%
\bibitem [{\citenamefont {Sharma}\ \emph {et~al.}(2004)\citenamefont {Sharma},
  \citenamefont {Ahn}, \citenamefont {Hur}, \citenamefont {Park}, \citenamefont
  {Kim}, \citenamefont {Lee}, \citenamefont {Park}, \citenamefont {Guha},\ and\
  \citenamefont {Cheong}}]{sharma2004thermal}%
  \BibitemOpen
  \bibfield  {author} {\bibinfo {author} {\bibfnamefont {P.~A.}\ \bibnamefont
  {Sharma}}, \bibinfo {author} {\bibfnamefont {J.}~\bibnamefont {Ahn}},
  \bibinfo {author} {\bibfnamefont {N.}~\bibnamefont {Hur}}, \bibinfo {author}
  {\bibfnamefont {S.}~\bibnamefont {Park}}, \bibinfo {author} {\bibfnamefont
  {S.~B.}\ \bibnamefont {Kim}}, \bibinfo {author} {\bibfnamefont
  {S.}~\bibnamefont {Lee}}, \bibinfo {author} {\bibfnamefont {J.-G.}\
  \bibnamefont {Park}}, \bibinfo {author} {\bibfnamefont {S.}~\bibnamefont
  {Guha}},\ and\ \bibinfo {author} {\bibfnamefont {S.-W.}\ \bibnamefont
  {Cheong}},\ }\bibfield  {title} {\bibinfo {title} {Thermal conductivity of
  geometrically frustrated, ferroelectric {YMnO$_3$}: Extraordinary spin-phonon
  interactions},\ }\href@noop {} {\bibfield  {journal} {\bibinfo  {journal}
  {Physical Review Letters}\ }\textbf {\bibinfo {volume} {93}},\ \bibinfo
  {pages} {177202} (\bibinfo {year} {2004})}\BibitemShut {NoStop}%
\bibitem [{\citenamefont {Kroumova}\ \emph {et~al.}(2003)\citenamefont
  {Kroumova}, \citenamefont {Aroyo}, \citenamefont {Perez-Mato}, \citenamefont
  {Kirov}, \citenamefont {Capillas}, \citenamefont {Ivantchev},\ and\
  \citenamefont {Wondratschek}}]{kroumova2003bilbao}%
  \BibitemOpen
  \bibfield  {author} {\bibinfo {author} {\bibfnamefont {E.}~\bibnamefont
  {Kroumova}}, \bibinfo {author} {\bibfnamefont {M.}~\bibnamefont {Aroyo}},
  \bibinfo {author} {\bibfnamefont {J.}~\bibnamefont {Perez-Mato}}, \bibinfo
  {author} {\bibfnamefont {A.}~\bibnamefont {Kirov}}, \bibinfo {author}
  {\bibfnamefont {C.}~\bibnamefont {Capillas}}, \bibinfo {author}
  {\bibfnamefont {S.}~\bibnamefont {Ivantchev}},\ and\ \bibinfo {author}
  {\bibfnamefont {H.}~\bibnamefont {Wondratschek}},\ }\bibfield  {title}
  {\bibinfo {title} {Bilbao crystallographic server: useful databases and tools
  for phase-transition studies},\ }\href@noop {} {\bibfield  {journal}
  {\bibinfo  {journal} {Phase Transitions: A Multinational Journal}\ }\textbf
  {\bibinfo {volume} {76}},\ \bibinfo {pages} {155} (\bibinfo {year}
  {2003})}\BibitemShut {NoStop}%
\bibitem [{\citenamefont {Iliev}\ \emph {et~al.}(2010)\citenamefont {Iliev},
  \citenamefont {Litvinchuk}, \citenamefont {Hadjiev}, \citenamefont
  {Gospodinov}, \citenamefont {Skumryev},\ and\ \citenamefont
  {Ressouche}}]{iliev2010phonon}%
  \BibitemOpen
  \bibfield  {author} {\bibinfo {author} {\bibfnamefont {M.}~\bibnamefont
  {Iliev}}, \bibinfo {author} {\bibfnamefont {A.}~\bibnamefont {Litvinchuk}},
  \bibinfo {author} {\bibfnamefont {V.}~\bibnamefont {Hadjiev}}, \bibinfo
  {author} {\bibfnamefont {M.}~\bibnamefont {Gospodinov}}, \bibinfo {author}
  {\bibfnamefont {V.}~\bibnamefont {Skumryev}},\ and\ \bibinfo {author}
  {\bibfnamefont {E.}~\bibnamefont {Ressouche}},\ }\bibfield  {title} {\bibinfo
  {title} {Phonon and magnon scattering of antiferromagnetic
  {Bi$_2$Fe$_4$O$_9$}},\ }\href@noop {} {\bibfield  {journal} {\bibinfo
  {journal} {Physical Review B}\ }\textbf {\bibinfo {volume} {81}},\ \bibinfo
  {pages} {024302} (\bibinfo {year} {2010})}\BibitemShut {NoStop}%
\bibitem [{\citenamefont {Kuroe}\ \emph {et~al.}(2008)\citenamefont {Kuroe},
  \citenamefont {Kusakabe}, \citenamefont {Oosawa}, \citenamefont {Sekine},
  \citenamefont {Yamada}, \citenamefont {Tanaka},\ and\ \citenamefont
  {Matsumoto}}]{kuroe2008magnetic}%
  \BibitemOpen
  \bibfield  {author} {\bibinfo {author} {\bibfnamefont {H.}~\bibnamefont
  {Kuroe}}, \bibinfo {author} {\bibfnamefont {K.}~\bibnamefont {Kusakabe}},
  \bibinfo {author} {\bibfnamefont {A.}~\bibnamefont {Oosawa}}, \bibinfo
  {author} {\bibfnamefont {T.}~\bibnamefont {Sekine}}, \bibinfo {author}
  {\bibfnamefont {F.}~\bibnamefont {Yamada}}, \bibinfo {author} {\bibfnamefont
  {H.}~\bibnamefont {Tanaka}},\ and\ \bibinfo {author} {\bibfnamefont
  {M.}~\bibnamefont {Matsumoto}},\ }\bibfield  {title} {\bibinfo {title}
  {Magnetic field induced one-magnon raman scattering in the magnon
  bose-einstein condensation phase of {TlCuCl$_3$}},\ }\href@noop {} {\bibfield
   {journal} {\bibinfo  {journal} {Physical Review B}\ }\textbf {\bibinfo
  {volume} {77}},\ \bibinfo {pages} {134420} (\bibinfo {year}
  {2008})}\BibitemShut {NoStop}%
\bibitem [{ame()}]{ament2010theory}%
  \BibitemOpen
  \bibfield  {title} {\bibinfo {title} {Theory of raman and resonant inelastic
  x-ray scattering from collective orbital excitations in {YTiO$_3$},
  author={Ament, LJP and Khaliullin, G}, journal={Physical Review B},
  volume={81}, number={12}, pages={125118}, year={2010}, publisher={APS}},\
  }\href@noop {} {\ }\BibitemShut {NoStop}%
\bibitem [{\citenamefont {Kr{\"u}ger}\ \emph {et~al.}(2004)\citenamefont
  {Kr{\"u}ger}, \citenamefont {Schulz}, \citenamefont {Naler}, \citenamefont
  {Rauer}, \citenamefont {Budelmann}, \citenamefont {B{\"a}ckstr{\"o}m},
  \citenamefont {Kim}, \citenamefont {Cheong}, \citenamefont {Perebeinos},\
  and\ \citenamefont {R{\"u}bhausen}}]{kruger2004orbital}%
  \BibitemOpen
  \bibfield  {author} {\bibinfo {author} {\bibfnamefont {R.}~\bibnamefont
  {Kr{\"u}ger}}, \bibinfo {author} {\bibfnamefont {B.}~\bibnamefont {Schulz}},
  \bibinfo {author} {\bibfnamefont {S.}~\bibnamefont {Naler}}, \bibinfo
  {author} {\bibfnamefont {R.}~\bibnamefont {Rauer}}, \bibinfo {author}
  {\bibfnamefont {D.}~\bibnamefont {Budelmann}}, \bibinfo {author}
  {\bibfnamefont {J.}~\bibnamefont {B{\"a}ckstr{\"o}m}}, \bibinfo {author}
  {\bibfnamefont {K.}~\bibnamefont {Kim}}, \bibinfo {author} {\bibfnamefont
  {S.}~\bibnamefont {Cheong}}, \bibinfo {author} {\bibfnamefont {.~f.~V.}\
  \bibnamefont {Perebeinos}},\ and\ \bibinfo {author} {\bibfnamefont
  {M.}~\bibnamefont {R{\"u}bhausen}},\ }\bibfield  {title} {\bibinfo {title}
  {Orbital ordering in {LaMnO$_3$} investigated by resonance raman
  spectroscopy},\ }\href@noop {} {\bibfield  {journal} {\bibinfo  {journal}
  {Physical Review Letters}\ }\textbf {\bibinfo {volume} {92}},\ \bibinfo
  {pages} {097203} (\bibinfo {year} {2004})}\BibitemShut {NoStop}%
\bibitem [{\citenamefont {Martin-Carron}\ and\ \citenamefont
  {De~Andr{\'e}s}(2004)}]{martin2004excitations}%
  \BibitemOpen
  \bibfield  {author} {\bibinfo {author} {\bibfnamefont {L.}~\bibnamefont
  {Martin-Carron}}\ and\ \bibinfo {author} {\bibfnamefont {A.}~\bibnamefont
  {De~Andr{\'e}s}},\ }\bibfield  {title} {\bibinfo {title} {Excitations of the
  orbital order in {RMnO$_3$} manganites: Light scattering experiments},\
  }\href@noop {} {\bibfield  {journal} {\bibinfo  {journal} {Physical Review
  Letters}\ }\textbf {\bibinfo {volume} {92}},\ \bibinfo {pages} {175501}
  (\bibinfo {year} {2004})}\BibitemShut {NoStop}%
\bibitem [{\citenamefont {Heyen}\ \emph {et~al.}(1992)\citenamefont {Heyen},
  \citenamefont {Kircher},\ and\ \citenamefont {Cardona}}]{heyen1992resonant}%
  \BibitemOpen
  \bibfield  {author} {\bibinfo {author} {\bibfnamefont {E.}~\bibnamefont
  {Heyen}}, \bibinfo {author} {\bibfnamefont {J.}~\bibnamefont {Kircher}},\
  and\ \bibinfo {author} {\bibfnamefont {M.}~\bibnamefont {Cardona}},\
  }\bibfield  {title} {\bibinfo {title} {Resonant raman scattering in
  insulating {YBa$_2$Cu$_3$O$_6$} as a probe of its electronic structure},\
  }\href@noop {} {\bibfield  {journal} {\bibinfo  {journal} {Physical Review
  B}\ }\textbf {\bibinfo {volume} {45}},\ \bibinfo {pages} {3037} (\bibinfo
  {year} {1992})}\BibitemShut {NoStop}%
\bibitem [{\citenamefont {Korotin}\ \emph {et~al.}(2015)\citenamefont
  {Korotin}, \citenamefont {Mazurenko}, \citenamefont {Anisimov},\ and\
  \citenamefont {Streltsov}}]{korotin2015calculation}%
  \BibitemOpen
  \bibfield  {author} {\bibinfo {author} {\bibfnamefont {D.~M.}\ \bibnamefont
  {Korotin}}, \bibinfo {author} {\bibfnamefont {V.~V.}\ \bibnamefont
  {Mazurenko}}, \bibinfo {author} {\bibfnamefont {V.~I.}\ \bibnamefont
  {Anisimov}},\ and\ \bibinfo {author} {\bibfnamefont {S.~V.}\ \bibnamefont
  {Streltsov}},\ }\bibfield  {title} {\bibinfo {title} {Calculation of exchange
  constants of the heisenberg model in plane-wave-based methods using the
  green's function approach},\ }\href
  {https://doi.org/10.1103/PhysRevB.91.224405} {\bibfield  {journal} {\bibinfo
  {journal} {Physical Review B}\ }\textbf {\bibinfo {volume} {91}},\ \bibinfo
  {pages} {224405} (\bibinfo {year} {2015})}\BibitemShut {NoStop}%
\bibitem [{\citenamefont {Balkanski}\ \emph {et~al.}(1983)\citenamefont
  {Balkanski}, \citenamefont {Wallis},\ and\ \citenamefont
  {Haro}}]{balkanski1983anharmonic}%
  \BibitemOpen
  \bibfield  {author} {\bibinfo {author} {\bibfnamefont {M.}~\bibnamefont
  {Balkanski}}, \bibinfo {author} {\bibfnamefont {R.}~\bibnamefont {Wallis}},\
  and\ \bibinfo {author} {\bibfnamefont {E.}~\bibnamefont {Haro}},\ }\bibfield
  {title} {\bibinfo {title} {Anharmonic effects in light scattering due to
  optical phonons in silicon},\ }\href@noop {} {\bibfield  {journal} {\bibinfo
  {journal} {Physical Review B}\ }\textbf {\bibinfo {volume} {28}},\ \bibinfo
  {pages} {1928} (\bibinfo {year} {1983})}\BibitemShut {NoStop}%
\bibitem [{\citenamefont {Kunwar}\ \emph {et~al.}(2024)\citenamefont {Kunwar},
  \citenamefont {Isha}, \citenamefont {Yogi}, \citenamefont {De}, \citenamefont
  {Dwij}, \citenamefont {Gupta}, \citenamefont {Mittal}, \citenamefont
  {Venkatesh}, \citenamefont {Chaudhary}, \citenamefont {Vedpathak} \emph
  {et~al.}}]{kunwar2024raman}%
  \BibitemOpen
  \bibfield  {author} {\bibinfo {author} {\bibfnamefont {H.~S.}\ \bibnamefont
  {Kunwar}}, \bibinfo {author} {\bibnamefont {Isha}}, \bibinfo {author}
  {\bibfnamefont {A.~K.}\ \bibnamefont {Yogi}}, \bibinfo {author}
  {\bibfnamefont {B.~K.}\ \bibnamefont {De}}, \bibinfo {author} {\bibfnamefont
  {V.}~\bibnamefont {Dwij}}, \bibinfo {author} {\bibfnamefont {M.~K.}\
  \bibnamefont {Gupta}}, \bibinfo {author} {\bibfnamefont {R.}~\bibnamefont
  {Mittal}}, \bibinfo {author} {\bibfnamefont {R.}~\bibnamefont {Venkatesh}},
  \bibinfo {author} {\bibfnamefont {R.}~\bibnamefont {Chaudhary}}, \bibinfo
  {author} {\bibfnamefont {M.}~\bibnamefont {Vedpathak}}, \emph {et~al.},\
  }\bibfield  {title} {\bibinfo {title} {Raman scattering of spin-1/2
  mixed-dimensionality antiferromagnetic $\alpha$-{Cu$_2$V$_2$O$_7$}},\
  }\href@noop {} {\bibfield  {journal} {\bibinfo  {journal} {Physical Review
  B}\ }\textbf {\bibinfo {volume} {109}},\ \bibinfo {pages} {054310} (\bibinfo
  {year} {2024})}\BibitemShut {NoStop}%
\bibitem [{\citenamefont {Badola}\ \emph {et~al.}(2024)\citenamefont {Badola},
  \citenamefont {Mukherjee}, \citenamefont {Sunil}, \citenamefont {Ghosh},
  \citenamefont {Negi}, \citenamefont {Vaitheeswaran}, \citenamefont
  {Garcia-Castro},\ and\ \citenamefont {Saha}}]{badola2024spin}%
  \BibitemOpen
  \bibfield  {author} {\bibinfo {author} {\bibfnamefont {S.}~\bibnamefont
  {Badola}}, \bibinfo {author} {\bibfnamefont {S.}~\bibnamefont {Mukherjee}},
  \bibinfo {author} {\bibfnamefont {G.}~\bibnamefont {Sunil}}, \bibinfo
  {author} {\bibfnamefont {B.}~\bibnamefont {Ghosh}}, \bibinfo {author}
  {\bibfnamefont {D.}~\bibnamefont {Negi}}, \bibinfo {author} {\bibfnamefont
  {G.}~\bibnamefont {Vaitheeswaran}}, \bibinfo {author} {\bibfnamefont
  {A.}~\bibnamefont {Garcia-Castro}},\ and\ \bibinfo {author} {\bibfnamefont
  {S.}~\bibnamefont {Saha}},\ }\bibfield  {title} {\bibinfo {title}
  {Spin-phonon coupling suppressing the structural transition in
  perovskite-like oxide},\ }\href@noop {} {\bibfield  {journal} {\bibinfo
  {journal} {Physical Review B}\ }\textbf {\bibinfo {volume} {109}},\ \bibinfo
  {pages} {L060104} (\bibinfo {year} {2024})}\BibitemShut {NoStop}%
\bibitem [{\citenamefont {Granado}\ \emph {et~al.}(1999)\citenamefont
  {Granado}, \citenamefont {Garc{\'\i}a}, \citenamefont {Sanjurjo},
  \citenamefont {Rettori}, \citenamefont {Torriani}, \citenamefont {Prado},
  \citenamefont {S{\'a}nchez}, \citenamefont {Caneiro},\ and\ \citenamefont
  {Oseroff}}]{granado1999magnetic}%
  \BibitemOpen
  \bibfield  {author} {\bibinfo {author} {\bibfnamefont {E.}~\bibnamefont
  {Granado}}, \bibinfo {author} {\bibfnamefont {A.}~\bibnamefont
  {Garc{\'\i}a}}, \bibinfo {author} {\bibfnamefont {J.}~\bibnamefont
  {Sanjurjo}}, \bibinfo {author} {\bibfnamefont {C.}~\bibnamefont {Rettori}},
  \bibinfo {author} {\bibfnamefont {I.}~\bibnamefont {Torriani}}, \bibinfo
  {author} {\bibfnamefont {F.}~\bibnamefont {Prado}}, \bibinfo {author}
  {\bibfnamefont {R.}~\bibnamefont {S{\'a}nchez}}, \bibinfo {author}
  {\bibfnamefont {A.}~\bibnamefont {Caneiro}},\ and\ \bibinfo {author}
  {\bibfnamefont {S.}~\bibnamefont {Oseroff}},\ }\bibfield  {title} {\bibinfo
  {title} {Magnetic ordering effects in the raman spectra of
  {$\mathrm{La_{1-x}Mn_{1-x}O_3}$}},\ }\href@noop {} {\bibfield  {journal}
  {\bibinfo  {journal} {Physical Review B}\ }\textbf {\bibinfo {volume} {60}},\
  \bibinfo {pages} {11879} (\bibinfo {year} {1999})}\BibitemShut {NoStop}%
\bibitem [{\citenamefont {Laverdi{\`e}re}\ \emph {et~al.}(2006)\citenamefont
  {Laverdi{\`e}re}, \citenamefont {Jandl}, \citenamefont {Mukhin},
  \citenamefont {Ivanov}, \citenamefont {Ivanov},\ and\ \citenamefont
  {Iliev}}]{laverdiere2006spin}%
  \BibitemOpen
  \bibfield  {author} {\bibinfo {author} {\bibfnamefont {J.}~\bibnamefont
  {Laverdi{\`e}re}}, \bibinfo {author} {\bibfnamefont {S.}~\bibnamefont
  {Jandl}}, \bibinfo {author} {\bibfnamefont {A.}~\bibnamefont {Mukhin}},
  \bibinfo {author} {\bibfnamefont {V.~Y.}\ \bibnamefont {Ivanov}}, \bibinfo
  {author} {\bibfnamefont {V.}~\bibnamefont {Ivanov}},\ and\ \bibinfo {author}
  {\bibfnamefont {M.}~\bibnamefont {Iliev}},\ }\bibfield  {title} {\bibinfo
  {title} {Spin-phonon coupling in orthorhombic {RMnO$_3$} {(R= Pr, Nd, Sm, Eu,
  Gd, Tb, Dy, Ho, Y)}: {A Raman study}},\ }\href@noop {} {\bibfield  {journal}
  {\bibinfo  {journal} {Physical Review B}\ }\textbf {\bibinfo {volume} {73}},\
  \bibinfo {pages} {214301} (\bibinfo {year} {2006})}\BibitemShut {NoStop}%
\bibitem [{\citenamefont {Cheong}\ and\ \citenamefont
  {Mostovoy}(2007)}]{cheong2007multiferroics}%
  \BibitemOpen
  \bibfield  {author} {\bibinfo {author} {\bibfnamefont {S.-W.}\ \bibnamefont
  {Cheong}}\ and\ \bibinfo {author} {\bibfnamefont {M.}~\bibnamefont
  {Mostovoy}},\ }\bibfield  {title} {\bibinfo {title} {Multiferroics: a
  magnetic twist for ferroelectricity},\ }\href@noop {} {\bibfield  {journal}
  {\bibinfo  {journal} {Nature Materials}\ }\textbf {\bibinfo {volume} {6}},\
  \bibinfo {pages} {13} (\bibinfo {year} {2007})}\BibitemShut {NoStop}%
\bibitem [{\citenamefont {Aguilar}\ \emph {et~al.}(2006)\citenamefont
  {Aguilar}, \citenamefont {Sushkov}, \citenamefont {Park}, \citenamefont
  {Cheong},\ and\ \citenamefont {Drew}}]{aguilar2006infrared}%
  \BibitemOpen
  \bibfield  {author} {\bibinfo {author} {\bibfnamefont {R.~V.}\ \bibnamefont
  {Aguilar}}, \bibinfo {author} {\bibfnamefont {A.}~\bibnamefont {Sushkov}},
  \bibinfo {author} {\bibfnamefont {S.}~\bibnamefont {Park}}, \bibinfo {author}
  {\bibfnamefont {S.-W.}\ \bibnamefont {Cheong}},\ and\ \bibinfo {author}
  {\bibfnamefont {H.}~\bibnamefont {Drew}},\ }\bibfield  {title} {\bibinfo
  {title} {Infrared phonon signatures of multiferroicity in {TbMn$_2$O$_5$}},\
  }\href@noop {} {\bibfield  {journal} {\bibinfo  {journal} {Physical Review
  B}\ }\textbf {\bibinfo {volume} {74}},\ \bibinfo {pages} {184404} (\bibinfo
  {year} {2006})}\BibitemShut {NoStop}%
\bibitem [{\citenamefont {Bal{\'e}dent}\ \emph {et~al.}(2015)\citenamefont
  {Bal{\'e}dent}, \citenamefont {Chattopadhyay}, \citenamefont {Fertey},
  \citenamefont {Lepetit}, \citenamefont {Greenblatt}, \citenamefont {Wanklyn},
  \citenamefont {Saouma}, \citenamefont {Jang},\ and\ \citenamefont
  {Foury-Leylekian}}]{baledent2015evidence}%
  \BibitemOpen
  \bibfield  {author} {\bibinfo {author} {\bibfnamefont {V.}~\bibnamefont
  {Bal{\'e}dent}}, \bibinfo {author} {\bibfnamefont {S.}~\bibnamefont
  {Chattopadhyay}}, \bibinfo {author} {\bibfnamefont {P.}~\bibnamefont
  {Fertey}}, \bibinfo {author} {\bibfnamefont {M.}~\bibnamefont {Lepetit}},
  \bibinfo {author} {\bibfnamefont {M.}~\bibnamefont {Greenblatt}}, \bibinfo
  {author} {\bibfnamefont {B.}~\bibnamefont {Wanklyn}}, \bibinfo {author}
  {\bibfnamefont {F.}~\bibnamefont {Saouma}}, \bibinfo {author} {\bibfnamefont
  {J.}~\bibnamefont {Jang}},\ and\ \bibinfo {author} {\bibfnamefont
  {P.}~\bibnamefont {Foury-Leylekian}},\ }\bibfield  {title} {\bibinfo {title}
  {Evidence for room temperature electric polarization in {RMn$_2$O$_5$}
  multiferroics},\ }\href@noop {} {\bibfield  {journal} {\bibinfo  {journal}
  {Physical Review Letters}\ }\textbf {\bibinfo {volume} {114}},\ \bibinfo
  {pages} {117601} (\bibinfo {year} {2015})}\BibitemShut {NoStop}%
\bibitem [{\citenamefont {Chattopadhyay}\ \emph {et~al.}(2017)\citenamefont
  {Chattopadhyay}, \citenamefont {Petit}, \citenamefont {Ressouche},
  \citenamefont {Raymond}, \citenamefont {Bal{\'e}dent}, \citenamefont {Yahia},
  \citenamefont {Peng}, \citenamefont {Robert}, \citenamefont {Lepetit},
  \citenamefont {Greenblatt} \emph {et~al.}}]{chattopadhyay20173}%
  \BibitemOpen
  \bibfield  {author} {\bibinfo {author} {\bibfnamefont {S.}~\bibnamefont
  {Chattopadhyay}}, \bibinfo {author} {\bibfnamefont {S.}~\bibnamefont
  {Petit}}, \bibinfo {author} {\bibfnamefont {E.}~\bibnamefont {Ressouche}},
  \bibinfo {author} {\bibfnamefont {S.}~\bibnamefont {Raymond}}, \bibinfo
  {author} {\bibfnamefont {V.}~\bibnamefont {Bal{\'e}dent}}, \bibinfo {author}
  {\bibfnamefont {G.}~\bibnamefont {Yahia}}, \bibinfo {author} {\bibfnamefont
  {W.}~\bibnamefont {Peng}}, \bibinfo {author} {\bibfnamefont {J.}~\bibnamefont
  {Robert}}, \bibinfo {author} {\bibfnamefont {M.-B.}\ \bibnamefont {Lepetit}},
  \bibinfo {author} {\bibfnamefont {M.}~\bibnamefont {Greenblatt}}, \emph
  {et~al.},\ }\bibfield  {title} {\bibinfo {title} {3d-4f coupling and
  multiferroicity in frustrated cairo pentagonal oxide {DyMn$_2$O$_5$}},\
  }\href@noop {} {\bibfield  {journal} {\bibinfo  {journal} {Scientific
  Reports}\ }\textbf {\bibinfo {volume} {7}},\ \bibinfo {pages} {14506}
  (\bibinfo {year} {2017})}\BibitemShut {NoStop}%
\bibitem [{\citenamefont {Garc{\'\i}a-Flores}\ \emph
  {et~al.}(2007)\citenamefont {Garc{\'\i}a-Flores}, \citenamefont {Granado},
  \citenamefont {Martinho}, \citenamefont {Rettori}, \citenamefont
  {Golovenchits}, \citenamefont {Sanina}, \citenamefont {Oseroff},
  \citenamefont {Park},\ and\ \citenamefont {Cheong}}]{garcia2007magnetically}%
  \BibitemOpen
  \bibfield  {author} {\bibinfo {author} {\bibfnamefont {A.}~\bibnamefont
  {Garc{\'\i}a-Flores}}, \bibinfo {author} {\bibfnamefont {E.}~\bibnamefont
  {Granado}}, \bibinfo {author} {\bibfnamefont {H.}~\bibnamefont {Martinho}},
  \bibinfo {author} {\bibfnamefont {C.}~\bibnamefont {Rettori}}, \bibinfo
  {author} {\bibfnamefont {E.}~\bibnamefont {Golovenchits}}, \bibinfo {author}
  {\bibfnamefont {V.}~\bibnamefont {Sanina}}, \bibinfo {author} {\bibfnamefont
  {S.}~\bibnamefont {Oseroff}}, \bibinfo {author} {\bibfnamefont
  {S.}~\bibnamefont {Park}},\ and\ \bibinfo {author} {\bibfnamefont {S.-W.}\
  \bibnamefont {Cheong}},\ }\bibfield  {title} {\bibinfo {title} {Magnetically
  frustrated behavior in multiferroics {RMn$_2$O$_5$} {(R= Bi, Eu, and Dy):} a
  {Raman} scattering study},\ }\href@noop {} {\bibfield  {journal} {\bibinfo
  {journal} {Journal of Applied Physics}\ }\textbf {\bibinfo {volume} {101}}
  (\bibinfo {year} {2007})}\BibitemShut {NoStop}%
\bibitem [{\citenamefont {Kagomiya}\ \emph {et~al.}(2003)\citenamefont
  {Kagomiya}, \citenamefont {Matsumoto}, \citenamefont {Kohn}, \citenamefont
  {Fukuda}, \citenamefont {Shoubu}, \citenamefont {Kimura}, \citenamefont
  {Noda},\ and\ \citenamefont {Ikeda}}]{kagomiya2003lattice}%
  \BibitemOpen
  \bibfield  {author} {\bibinfo {author} {\bibfnamefont {I.}~\bibnamefont
  {Kagomiya}}, \bibinfo {author} {\bibfnamefont {S.}~\bibnamefont {Matsumoto}},
  \bibinfo {author} {\bibfnamefont {K.}~\bibnamefont {Kohn}}, \bibinfo {author}
  {\bibfnamefont {Y.}~\bibnamefont {Fukuda}}, \bibinfo {author} {\bibfnamefont
  {T.}~\bibnamefont {Shoubu}}, \bibinfo {author} {\bibfnamefont
  {H.}~\bibnamefont {Kimura}}, \bibinfo {author} {\bibfnamefont
  {Y.}~\bibnamefont {Noda}},\ and\ \bibinfo {author} {\bibfnamefont
  {N.}~\bibnamefont {Ikeda}},\ }\bibfield  {title} {\bibinfo {title} {Lattice
  distortion at ferroelectric transition of {YMn$_2$O$_5$}},\ }\href@noop {}
  {\bibfield  {journal} {\bibinfo  {journal} {Ferroelectrics}\ }\textbf
  {\bibinfo {volume} {286}},\ \bibinfo {pages} {167} (\bibinfo {year}
  {2003})}\BibitemShut {NoStop}%
\bibitem [{\citenamefont {Kim}\ \emph {et~al.}(2011)\citenamefont {Kim},
  \citenamefont {Van Der~Vegte}, \citenamefont {Scaramucci}, \citenamefont
  {Artyukhin}, \citenamefont {Chung}, \citenamefont {Park}, \citenamefont
  {Cheong}, \citenamefont {Mostovoy},\ and\ \citenamefont
  {Lee}}]{kim2011magnetic}%
  \BibitemOpen
  \bibfield  {author} {\bibinfo {author} {\bibfnamefont {J.-H.}\ \bibnamefont
  {Kim}}, \bibinfo {author} {\bibfnamefont {M.}~\bibnamefont {Van Der~Vegte}},
  \bibinfo {author} {\bibfnamefont {A.}~\bibnamefont {Scaramucci}}, \bibinfo
  {author} {\bibfnamefont {S.}~\bibnamefont {Artyukhin}}, \bibinfo {author}
  {\bibfnamefont {J.-H.}\ \bibnamefont {Chung}}, \bibinfo {author}
  {\bibfnamefont {S.}~\bibnamefont {Park}}, \bibinfo {author} {\bibfnamefont
  {S.}~\bibnamefont {Cheong}}, \bibinfo {author} {\bibfnamefont
  {M.}~\bibnamefont {Mostovoy}},\ and\ \bibinfo {author} {\bibfnamefont
  {S.-H.}\ \bibnamefont {Lee}},\ }\bibfield  {title} {\bibinfo {title}
  {Magnetic excitations in the low-temperature ferroelectric phase of
  multiferroic {YMn$_2$O$_5$} using inelastic neutron scattering},\ }\href@noop
  {} {\bibfield  {journal} {\bibinfo  {journal} {Physical Review Letters}\
  }\textbf {\bibinfo {volume} {107}},\ \bibinfo {pages} {097401} (\bibinfo
  {year} {2011})}\BibitemShut {NoStop}%
\bibitem [{\citenamefont {Zhao}\ \emph {et~al.}(2014)\citenamefont {Zhao},
  \citenamefont {Liu}, \citenamefont {Li}, \citenamefont {Lin}, \citenamefont
  {Yan}, \citenamefont {Dong},\ and\ \citenamefont
  {Liu}}]{zhao2014experimental}%
  \BibitemOpen
  \bibfield  {author} {\bibinfo {author} {\bibfnamefont {Z.}~\bibnamefont
  {Zhao}}, \bibinfo {author} {\bibfnamefont {M.}~\bibnamefont {Liu}}, \bibinfo
  {author} {\bibfnamefont {X.}~\bibnamefont {Li}}, \bibinfo {author}
  {\bibfnamefont {L.}~\bibnamefont {Lin}}, \bibinfo {author} {\bibfnamefont
  {Z.}~\bibnamefont {Yan}}, \bibinfo {author} {\bibfnamefont {S.}~\bibnamefont
  {Dong}},\ and\ \bibinfo {author} {\bibfnamefont {J.-M.}\ \bibnamefont
  {Liu}},\ }\bibfield  {title} {\bibinfo {title} {Experimental observation of
  ferrielectricity in multiferroic {DyMn$_2$O$_5$}},\ }\href@noop {} {\bibfield
   {journal} {\bibinfo  {journal} {Scientific Reports}\ }\textbf {\bibinfo
  {volume} {4}},\ \bibinfo {pages} {3984} (\bibinfo {year} {2014})}\BibitemShut
  {NoStop}%
\bibitem [{\citenamefont {Mihailova}\ \emph {et~al.}(2005)\citenamefont
  {Mihailova}, \citenamefont {Gospodinov}, \citenamefont {G{\"u}ttler},
  \citenamefont {Yen}, \citenamefont {Litvinchuk},\ and\ \citenamefont
  {Iliev}}]{mihailova2005temperature}%
  \BibitemOpen
  \bibfield  {author} {\bibinfo {author} {\bibfnamefont {B.}~\bibnamefont
  {Mihailova}}, \bibinfo {author} {\bibfnamefont {M.}~\bibnamefont
  {Gospodinov}}, \bibinfo {author} {\bibfnamefont {B.}~\bibnamefont
  {G{\"u}ttler}}, \bibinfo {author} {\bibfnamefont {F.}~\bibnamefont {Yen}},
  \bibinfo {author} {\bibfnamefont {A.}~\bibnamefont {Litvinchuk}},\ and\
  \bibinfo {author} {\bibfnamefont {M.}~\bibnamefont {Iliev}},\ }\bibfield
  {title} {\bibinfo {title} {Temperature-dependent raman spectra of
  {HoMn$_2$O$_5$} and {TbMn$_2$O$_5$}},\ }\href@noop {} {\bibfield  {journal}
  {\bibinfo  {journal} {Physical Review B}\ }\textbf {\bibinfo {volume} {71}},\
  \bibinfo {pages} {172301} (\bibinfo {year} {2005})}\BibitemShut {NoStop}%
\bibitem [{\citenamefont {Chapon}\ \emph {et~al.}(2004)\citenamefont {Chapon},
  \citenamefont {Blake}, \citenamefont {Gutmann}, \citenamefont {Park},
  \citenamefont {Hur}, \citenamefont {Radaelli},\ and\ \citenamefont
  {Cheong}}]{chapon2004structural}%
  \BibitemOpen
  \bibfield  {author} {\bibinfo {author} {\bibfnamefont {L.}~\bibnamefont
  {Chapon}}, \bibinfo {author} {\bibfnamefont {G.}~\bibnamefont {Blake}},
  \bibinfo {author} {\bibfnamefont {M.~J.}\ \bibnamefont {Gutmann}}, \bibinfo
  {author} {\bibfnamefont {S.}~\bibnamefont {Park}}, \bibinfo {author}
  {\bibfnamefont {N.}~\bibnamefont {Hur}}, \bibinfo {author} {\bibfnamefont
  {P.~G.}\ \bibnamefont {Radaelli}},\ and\ \bibinfo {author} {\bibfnamefont
  {S.-W.}\ \bibnamefont {Cheong}},\ }\bibfield  {title} {\bibinfo {title}
  {Structural anomalies and multiferroic behavior in magnetically frustrated
  {TbMn$_2$O$_5$}},\ }\href@noop {} {\bibfield  {journal} {\bibinfo  {journal}
  {Physical Review Letters}\ }\textbf {\bibinfo {volume} {93}},\ \bibinfo
  {pages} {177402} (\bibinfo {year} {2004})}\BibitemShut {NoStop}%
\bibitem [{\citenamefont {Park}\ \emph {et~al.}(2010)\citenamefont {Park},
  \citenamefont {Song}, \citenamefont {Lee}, \citenamefont {Won},\ and\
  \citenamefont {Hur}}]{park2010effect}%
  \BibitemOpen
  \bibfield  {author} {\bibinfo {author} {\bibfnamefont {Y.}~\bibnamefont
  {Park}}, \bibinfo {author} {\bibfnamefont {K.}~\bibnamefont {Song}}, \bibinfo
  {author} {\bibfnamefont {K.}~\bibnamefont {Lee}}, \bibinfo {author}
  {\bibfnamefont {C.}~\bibnamefont {Won}},\ and\ \bibinfo {author}
  {\bibfnamefont {N.}~\bibnamefont {Hur}},\ }\bibfield  {title} {\bibinfo
  {title} {Effect of antiferromagnetic order on the dielectric properties of
  {Bi$_2$Fe$_4$O$_9$}},\ }\href@noop {} {\bibfield  {journal} {\bibinfo
  {journal} {Applied Physics Letters}\ }\textbf {\bibinfo {volume} {96}}
  (\bibinfo {year} {2010})}\BibitemShut {NoStop}%
\bibitem [{\citenamefont {Stockem}\ \emph {et~al.}(2018)\citenamefont
  {Stockem}, \citenamefont {Bergman}, \citenamefont {Glensk}, \citenamefont
  {Hickel}, \citenamefont {K{\"o}rmann}, \citenamefont {Grabowski},
  \citenamefont {Neugebauer},\ and\ \citenamefont
  {Alling}}]{stockem2018anomalous}%
  \BibitemOpen
  \bibfield  {author} {\bibinfo {author} {\bibfnamefont {I.}~\bibnamefont
  {Stockem}}, \bibinfo {author} {\bibfnamefont {A.}~\bibnamefont {Bergman}},
  \bibinfo {author} {\bibfnamefont {A.}~\bibnamefont {Glensk}}, \bibinfo
  {author} {\bibfnamefont {T.}~\bibnamefont {Hickel}}, \bibinfo {author}
  {\bibfnamefont {F.}~\bibnamefont {K{\"o}rmann}}, \bibinfo {author}
  {\bibfnamefont {B.}~\bibnamefont {Grabowski}}, \bibinfo {author}
  {\bibfnamefont {J.}~\bibnamefont {Neugebauer}},\ and\ \bibinfo {author}
  {\bibfnamefont {B.}~\bibnamefont {Alling}},\ }\bibfield  {title} {\bibinfo
  {title} {Anomalous phonon lifetime shortening in paramagnetic {CrN} caused by
  spin-lattice coupling: a combined spin and $ab-initio$ molecular dynamics
  study},\ }\href@noop {} {\bibfield  {journal} {\bibinfo  {journal} {Physical
  Review Letters}\ }\textbf {\bibinfo {volume} {121}},\ \bibinfo {pages}
  {125902} (\bibinfo {year} {2018})}\BibitemShut {NoStop}%
\bibitem [{\citenamefont {Krenzel}\ \emph {et~al.}(2012)\citenamefont
  {Krenzel}, \citenamefont {Schreuer}, \citenamefont {Gesing}, \citenamefont
  {Burianek}, \citenamefont {M{\"u}hlberg},\ and\ \citenamefont
  {Schneider}}]{krenzel2012thermal}%
  \BibitemOpen
  \bibfield  {author} {\bibinfo {author} {\bibfnamefont {T.~F.}\ \bibnamefont
  {Krenzel}}, \bibinfo {author} {\bibfnamefont {J.}~\bibnamefont {Schreuer}},
  \bibinfo {author} {\bibfnamefont {T.~M.}\ \bibnamefont {Gesing}}, \bibinfo
  {author} {\bibfnamefont {M.}~\bibnamefont {Burianek}}, \bibinfo {author}
  {\bibfnamefont {M.}~\bibnamefont {M{\"u}hlberg}},\ and\ \bibinfo {author}
  {\bibfnamefont {H.}~\bibnamefont {Schneider}},\ }\bibfield  {title} {\bibinfo
  {title} {Thermal expansion and elastic properties of mullite-type
  {Bi$_2$Ga$_4$O$_9$} and {Bi$_2$Fe$_4$O$_9$} single crystals},\ }\href@noop {}
  {\bibfield  {journal} {\bibinfo  {journal} {International Journal of
  Materials Research}\ }\textbf {\bibinfo {volume} {103}},\ \bibinfo {pages}
  {438} (\bibinfo {year} {2012})}\BibitemShut {NoStop}%
\end{thebibliography}

%

	\pagebreak
	
\end{document}